\documentclass[twocolumn]{aastex62}
\usepackage[utf8]{inputenc}

\usepackage{natbib}
\usepackage{graphicx}
\usepackage{xcolor}
\usepackage{hyperref}
\usepackage{multirow}
\usepackage{amsmath}

\begin{document}

\title{Terminus: A Versatile Simulator for Space-based Telescopes}

\shorttitle{Terminus}
\shortauthors{Edwards \& Stotesbury}

\correspondingauthor{Billy Edwards}
\email{billy@bssl.space}
\author[0000-0002-5494-3237]{Billy Edwards}
\affiliation{Blue Skies Space Ltd., 69 Wilson Street, London, EC2A 2BB, UK}
\affiliation{Department of Physics and Astronomy, University College London, Gower Street, London, WC1E 6BT, UK}

\author{Ian Stotesbury}
\affiliation{Blue Skies Space Ltd., 69 Wilson Street, London, EC2A 2BB, UK}

\begin{abstract}

Space-based telescopes offer unparalleled opportunities for characterising exoplanets, Solar System bodies and stellar objects. However, observatories in low Earth orbits (e.g. Hubble, CHEOPS, Twinkle and an ever increasing number of cubesats) cannot always be continuously pointed at a target due to Earth obscuration. For exoplanet observations consisting of transit, or eclipse, spectroscopy this causes gaps in the light curve, which reduces the information content and can diminish the science return of the observation. Terminus, a time-domain simulator, has been developed to model the occurrence of these gaps to predict the potential impact on future observations. The simulator is capable of radiometrically modelling exoplanet observations as well as producing light curves and spectra. Here, Terminus is baselined on the Twinkle mission but the model can be adapted for any space-based telescope and is especially applicable to those in a low-Earth orbit. Terminus also has the capability to model observations of other targets such as asteroids or brown dwarfs. \\\\

\end{abstract}

\section{Introduction}

To date, several thousand extra-solar planets have been discovered. With many of these now being detected around bright stars, and with many more to come from missions such as the Transiting Exoplanet Survey Satellite (TESS, \cite{ricker,barclay}), the characterisation of these worlds has begun and will accelerate over the next decade. Ground-based instruments have detected absorption and emission lines in exoplanet atmospheres via high resolution spectra \citep[e.g.][]{hoeijmakers,ehrenreich_wasp76}) while the Hubble and Spitzer space telescopes have used lower resolution spectroscopy or photometry to probe the chemical abundances and thermal properties of tens of planets \citep[e.g.][]{sing,iyer_pop,tsiaras_30planets,garhart_spitzer}. 

In the coming years several missions, some of which are specifically designed for exoplanet research, will be launched to provide further characterisation. While the James Webb Space Telescope (JWST, \cite{greene}) and Ariel \citep{tinetti_ariel} will be located at L2, observatories such as the CHaracterising ExOPlanets Satellite (CHEOPS, \cite{benz_cheops}), which was launched in December 2019, and Twinkle \citep{edwards_exo} will operate from a low Earth orbit and as such will have to contend with Earth obscuration. 

The orbit will cause gaps in some of the observations obtained by these missions which will impact their information content due to parts of the transit light curve being missed, decreasing the precision of the recovered transit parameters. Additionally, the thermal environment of a low Earth orbit and the breaks in observing can lead to recurring systematic trends such as ramps in the recorded flux due to thermal breathing of the telescope and detector persistence. Such gaps and systematics are experienced in all exoplanet observations with Hubble \cite[e.g.][]{deming_hd209,kreidberg_gj1214}. It should be noted, however, that Hubble is situated in an equatorial orbit which is significantly different to the sun-synchronous orbits of CHEOPS and Twinkle. Sun-synchronous orbits allow for certain areas of sky, specifically those in the anti-sun direction, to be observed for longer periods without interruption. Additionally, the thermal environment is more stable due to the smaller variations in the spacecraft-Earth-Sun geometry. Previous missions to have operated in sun-synchronous orbits include the Convection, Rotation and planetary Transits (CoRoT, \citet{borde_corot}), Akari \citep{murakami_akari} and WISE/NEOWISE \citep{wright_wise,mainzer_neowise}. Due to it's Earth-trailing orbit, Spitzer \citep{werner_spitzer} did not experience gaps in its observations. 

When designing future instrumentation, understanding the expected performance for the envisioned science cases is paramount. Static models, often referred to as radiometric or sensitivity models, are suitable for studying the instrument performance over a wide parameter space (i.e. for many different targets) as they are generally quick to run and require relatively minimal information about the instrumentation. Radiometric models are a useful way to understand the capabilities of upcoming exoplanet observatories and have been widely used. The ESA Radiometric Model (ERM, \citet{puig}) was used to simulate the performance of the ESA M3 candidate EChO (Exoplanet Characterisation Observatory, \citet{tinetti_echo}) and was subsequently used for Ariel \citep{puig_2018}. A newer, python-based version, ArielRad, was recently developed \citep{mugnai_arielrad} while PandExo has been created for simulating exoplanet observations with Hubble and JWST \citep{pandexo} and the NIRSpec Exoplanet Exposure Time Calculator (NEETC) was built specifically for modelling transit and eclipse spectroscopy with JWST's NIRSpec instrument \citep{nielsen_neetc}. These usually account for efficiency of the optics and simple noise contributions such as photon, dark current, readout and instrument/telescope emission.

More complex effects, such as jitter, stellar variability and spots and correlated noise sources require models which have a time-domain aspect. These tools usually also produce simulated detector images which can act as realistic data products for the mission, accounting for detector effects such as correlated noise between pixels or inter- and intra-pixel variations. For example, ExoSim is a numerical end-to-end simulator of transit spectroscopy which is currently being utilised for the Ariel mission \citep{pascale_echosim,ExoSim,sarkar}. The tool has been created to explore a variety of signal and noise issues that occur in, and may bias, transit spectroscopy observations, including instrument systematics and the other effects previously mentioned. By producing realistic raw data products, the outputs can also be fed into data reduction pipelines to explore, and remove, potential biases within them as well as develop new reduction and data correction methods. End-to-end simulators such as ExoSim are therefore powerful tools for understanding the capabilities of an instrument design. Additional time-domain simulators of note include ExoNoodle \citep{exonoodle}, which utilises MIRISim \citep{geers_mirisim} to model time-series with the JWST MIRI instrument, Wayne which models Hubble spatial scans of exoplanets \citep{varley_wayne} and the simulators developed for the CHEOPS and Colorado Ultraviolet Transit Experiment (CUTE) missions \citep{futyan_cheops,sreejith_cute}. While the complexity of these types of tools can be hugely advantageous in understanding intricate effects it can also be their biggest weakness; such sophisticated models require a great deal of time to develop and run as well as an excellent understanding of all parts of the instrument design. They can therefore only be applied to highly refined designs and run for a small number of cases. The solution to the issue of complexity versus efficiency is to use both types of models. For Ariel, ExoSim is used to validate the outcomes of ArielRad for selected, representative targets. ArielRad is then used as the workhorse for modelling the capability of thousands of targets due to its superior speed \citep{edwards_ariel,mugnai_arielrad}.

Here, we describe the Terminus tool which has been developed to model transit (and eclipse) observations with Twinkle, to explore the impact of Earth obscuration and allow for efficient scheduling methods to be developed to minimise this impact. The simulator, however, is not mission specific and could be adapted for other observatories, with a particular applicability for satellites in low Earth orbit.

The Twinkle Space Mission\footnote{\url{http://www.twinkle-spacemission.co.uk}} is a new, fast-track satellite designed to begin science operations in 2024. It has been conceived for providing faster access to spectroscopic data from exoplanet atmospheres and Solar System bodies. Twinkle is equipped with a visible and infrared spectrometer which simultaneously covers 0.5-4.5 $\mu$m with a resolving power of R$\sim$20-70 across this range. Twinkle has been designed with a telescope aperture of 0.45 m. Twinkle's field of regard is a cone with an opening angle of 40$^\circ$, centred on the anti-sun vector \citep{savini}.

Previously the ESA Radiometric Model (ERM, \cite{puig,puig_2018}), which assumes full light curves are observed, has been used to model the capabilities of Twinkle (see \cite{edwards_exo}). Terminus includes a radiometric model, built upon the concepts of the ERM, but it has been upgraded to also have the capacity to simulate light curves. The code also contains the ability to model the orbit of a spacecraft, thus allowing for the availability of targets to be understood given solar, lunar and Earth exclusion angles. The capability to model these gaps is not available in other tools such as ArielRad or ExoSim and is one of the driving factors behind the creation of Terminus. Additionally, the Twinkle mission will not be limited to exoplanet characterisation and will also observe solar system bodies, brown dwarfs and other astrophysical objects. As such, Terminus builds upon the work of \cite{edwards_ast,edwards_solar} and can be used to calculate the predicted data quality and observational periods for these objects, another feature which is not present in other similar codes.

In this work we first describe the portion of the simulator which calculates the target signal and noise contributions before comparing the outputs of simulated light curve fitting to radiometric estimates. Next the orbital module is detailed and validated against outputs from an orbital dynamics software. Using this we explore the effect of gaps for observations of HD\,209458\,b and WASP-127\,b with Twinkle. Finally, we discuss Twinkle's ability to observe asteroids by focusing on potential observations of the Near-Earth Object (NEO) 99942 Apophis (2004 MN4).

\section{Simulator Structure}

\begin{figure}
    \centering
    \includegraphics[width=\columnwidth]{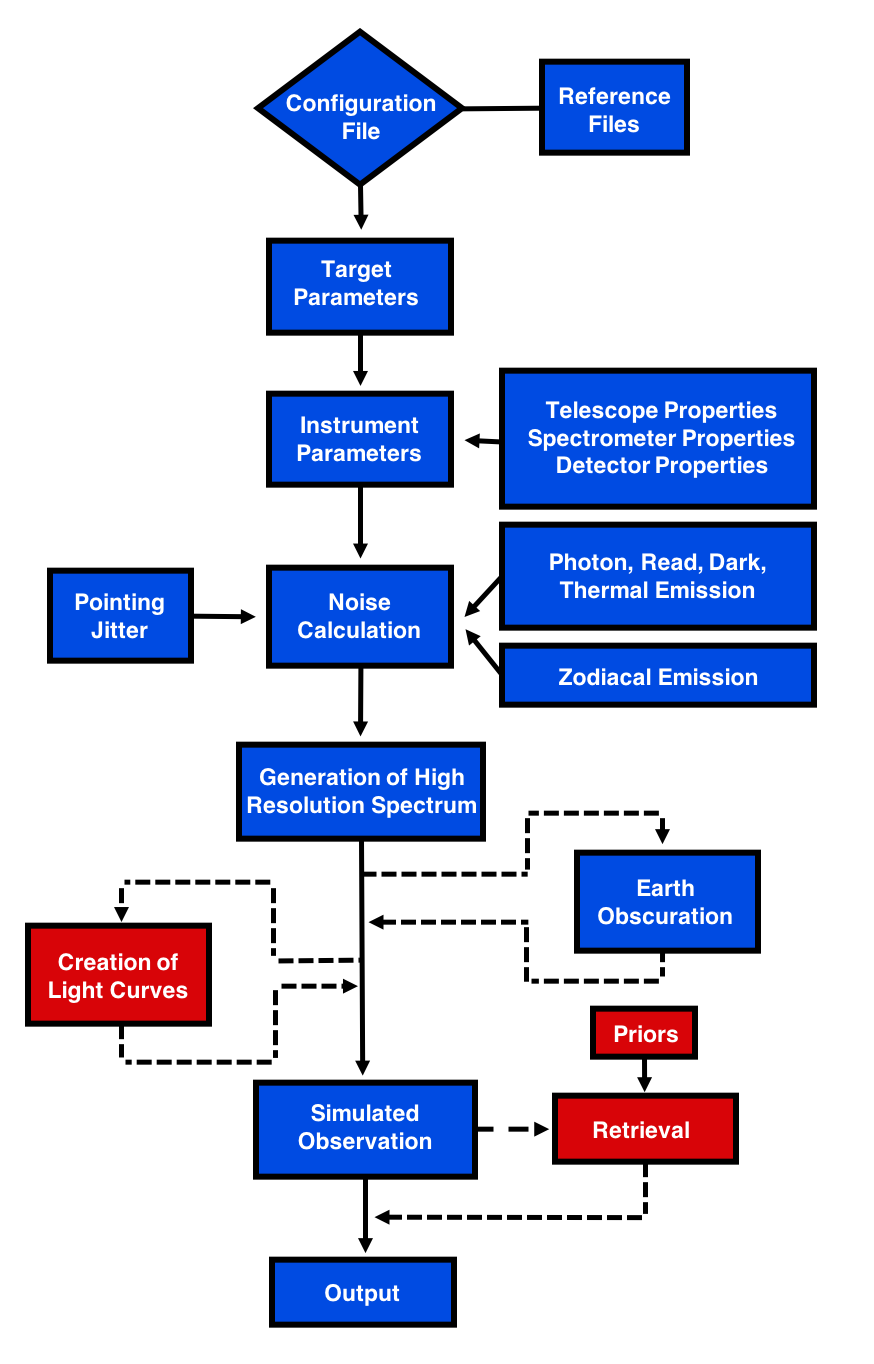}
    \caption{Overview of the simulator structure. Generic parts are represented by blue shapes while red indicates functions which are exoplanet specific. Dotted lines indicate portions which are not compulsory.}
    \label{fig:Terminus}
\end{figure}

Terminus has been constructed in Python and has several different stages. It can be operated as a simple radiometric model, used to calculate expected signal-to-noise ratio (SNR) on a given number of atmospheric scale heights, or be utilised to create simulated light curves. A instrument file is loaded (which includes parameters such as telescope aperture, quantum efficiency etc.) and the star flux on the detector calculated. PSFs can be imported from external sources. In Sections \ref{target_para} to \ref{lc_modelling} we discuss the structure of the simulator and an overview is given in Figure \ref{fig:Terminus}. 
\subsection{Target Parameters}
\label{target_para}

A catalogue of planets has been created following the methodology of \citep{edwards_ariel} and data is taken from the NASA Exoplanet Archive \citep{akeson_nasa_archive}.


\subsection{Radiometric Model}
\label{radiometric}

\begin{figure}
    \centering
    \includegraphics[width=\columnwidth]{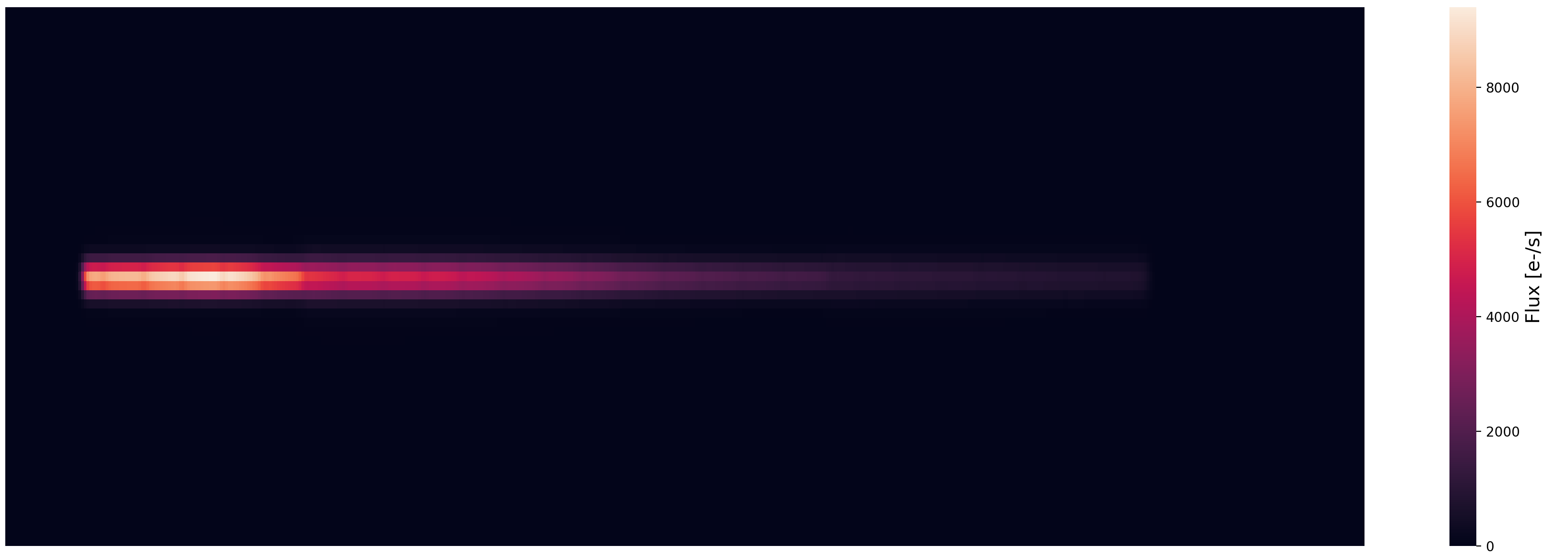}
    \includegraphics[width=\columnwidth]{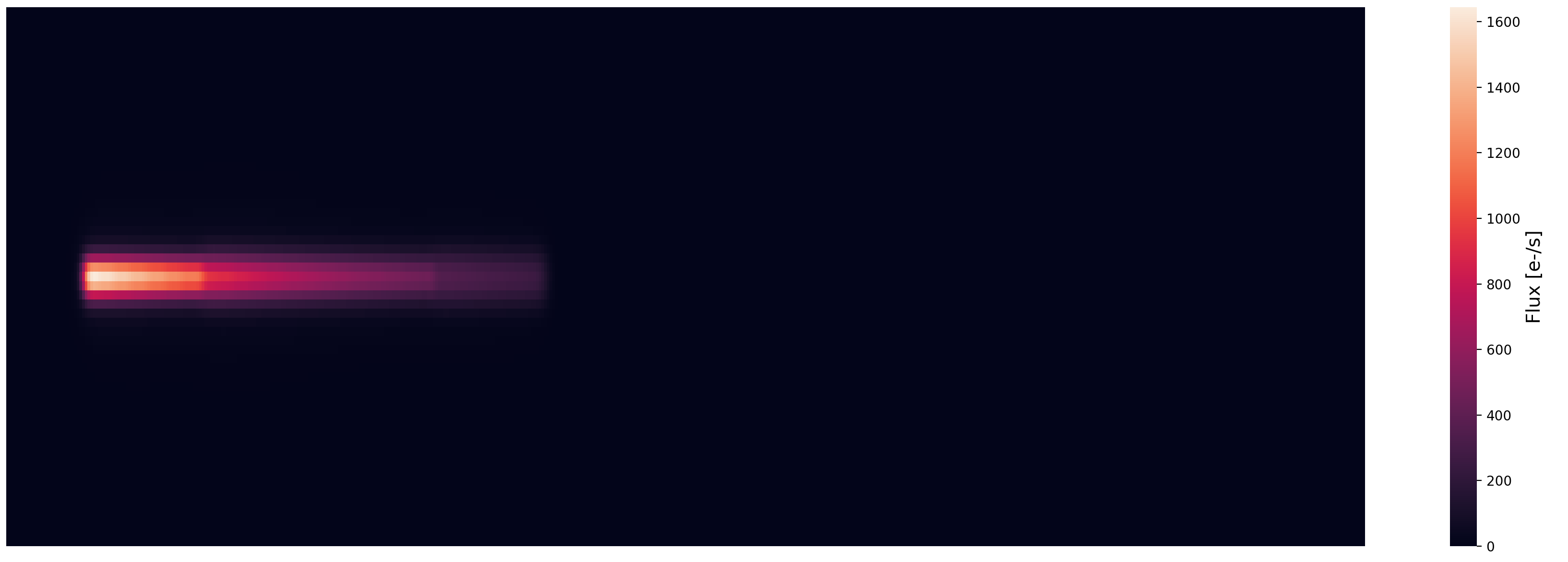}
    \caption{Example detector images generated by Terminus for Twinkle Ch0 (top) and Ch1 (bottom). These are used purely for the calculation of the saturation time for each target.}
    \label{fig:detector}
\end{figure}

The stellar flux at Earth is calculated using spectral energy distributions (SEDs) from the PHOENIX BT-Settl models by \cite{allard_phoenix, husser_phoenix, baraffe}. The spectral irradiance from a host star at the aperture of the telescope is given by:
\begin{equation}
E_S(\lambda) = S_S(\lambda)(\frac{R_*}{d})^2
\end{equation}
where S$_{\rm S}(\lambda)$ is the star spectral irradiance from the Phoenix catalogue (Wm$^{-2}\mu$m$^{-1}$) and d is the distance to the star. The effective collecting area of the telescope is then accounted for before the flux is integrated into the spectral bins of the instrumentation to give a photon flux per bin. The signal is then propagated through the instrument to the detector focal planes, taking into account the transmission of each optical component and diffracting element as well as the quantum efficiency of the detector. The final signal, in electrons per second, from the star in each spectral bin is determined as a 1D flux rate before being convolved with 2D point spread functions (PSFs) and the instrument dispersion to create a detector image. The detector image, like the one shown in Figure \ref{fig:detector}, is utilised to calculate the saturation time for the target while the 1D flux rate is used for all other calculations. A variety of sources of noise are accounted for in each of the models. In addition to photon noise, the simulator calculates the contributions from dark current, instrument and telescope emission, zodiacal background emission, and readout noise. Additionally, photometric uncertainties due to spacecraft jitter can be imported and interpolated from time-domain simulators such as ExoSim (see Section \ref{sec:jitter}). Some of these noise sources are wavelength dependent (e.g. zodiacal background) while others are not (e.g. read noise).

\subsubsection{Calculating Noise Per Exposure}

In describing the acquisition of data we use the nomenclature of \citet{rauscher} in which a frame refers to the clocking and digitisation of pixels within a specified area of the detector known as a sub-array. The size of sub-array dictates the time required for it to read out. Here, given the footprint of Twinkle's spectrometer on the detector, we assume a fastest frame time of 0.5 seconds which is similar to that for the 1024A sub-array on JWST NIRSpec (0.45 seconds, \cite{pandeia}). A collection of frames then forms a group although here, as with JWST time series, the number of frames is set to one (i.e. t$_{\rm g}$ = t$_{\rm f}$). A collection of non-destructively read groups, along with a detector reset, forms an integration. Here, the detector reset time after a destructive read is also assumed to be equivalent to the frame time. As the duration of a transit/eclipse is generally orders of magnitude longer than the saturation time of the detector, many integrations will be taken during an observation. The total noise variance per integration, $\sigma^2_{\rm exp}$, is given by:
\begin{equation}
    \sigma^2_{exp} = \frac{12(n_g-1)}{n_g(n_g+1)}n_{pix}\sigma^2_{read} + \frac{6(n_g^2 + 1)}{5n_g(n_g+1)}(n_g-1)t_gi_{total}
    \label{exp_noise}
\end{equation}
from \cite{rauscher} where n$_{\rm g}$ is the number of groups (non-destructive reads) per exposure, $\sigma_{\rm read}$ is the read noise in e$^-$/pix rms, n$_{\rm pix}$ is the number of pixels in the spectral bin, t$_{\rm g}$ is the time for a single non-destructive group read, and i$_{\rm total}$ is the total flux in e$^-$/s. For JWST observations, the standard practise for exoplanet observations is to maximise the number of groups \citep{pandexo}. Meanwhile, Ariel will use a variety of readout modes, depending upon the brightness of the target, with correlated double sampling (CDS, n$_{\rm g}$ =2) for brighter sources targets and multiple up-the-ramp reads for fainter targets \citep{focardi_ariel}. Collecting several up-the-ramp reads can be useful in correcting for cosmic ray impacts while also reducing the read noise. Additional reads, however, increase the photon noise contribution and thus Terminus varies the number of up-the-ramp reads according to the brightness of the target to attempt to optimise noise. In each case, the maximum number of up-the-ramp reads is calculated and Equation 2 used to selected the number of reads which yields the lowest noise per transit observation (using Equations 2-7). n$_{\rm pix}$ can be selected by specifying a required encircled energy but when importing jitter simulations from ExoSim, n$_{\rm pix}$ is set to the values used in these simulations as outlined in Section \ref{sec:jitter}. In Equation 2, i$_{\rm total}$ is defined as:
\begin{equation}
    i_{total} = i_{sig} + n_{pix}(i_{dark} + i_{bdg})
\end{equation}
where i$_{\rm sig}$ is the total signal from the star in the spectral bin (e$^-$/s) while i$_{\rm dark}$ and i$_{\rm bdg}$ are the dark current and background signals respectively (in e$^-$/s/pix). Currently, $i_{\rm bdg}$ is assumed to be from the emission of optical elements and Zodiacal emission, as detailed in Section \ref{sec:zodi}, but future updates will include contributions from nearby stars. For exoplanet spectroscopy, the total observational time is generally quantised in terms of the duration of a transit/eclipse event, T. The model assumes the time spent during ingress (T$_{\rm 12}$) and egress (T$_{\rm 34}$) is negligible to the primary transit time (T$_{23}$) and thus T = T$_{\rm 23}$ = T$_{\rm 14}$. The transit time can be calculated from:
\begin{equation}
	T_{14} = \sqrt{1-b^2}  \frac{R_* P}{\pi a}
\end{equation}
for a given system where P is the orbital period. The fractional noise on the star signal over one transit duration is then given by:
\begin{equation}
    \sigma_{Star} = \frac{1}{\sqrt{n_{int}}}\frac{\sigma_{exp}}{i_{sig}}
\end{equation}
where n$_{\rm int}$ is the number of integrations over one transit duration which is calculated from:
\begin{equation}
n_{int} = \frac{T_{14}}{t_r + t_g n_g}
\end{equation}
where t$_{\rm r}$ is the time taken to reset the detector. As a baseline we take t$_{\rm r}$ to be equivalent to the frame time, t$_{\rm f}$ (0.5 seconds). As noted by \cite{pandexo}, if 
t$_{\rm g}$ = t$_{\rm r}$ = t$_{\rm f}$ then the duty cycle (i.e. the efficiency) is given by (n$_{\rm g}$ $-$ 1)/(n$_{\rm g}$ $+$ 1).

The measurement of the transit depth is differential and thus the error (i.e. the 1$\sigma$ uncertainty) on the transit depth is given by:
\begin{equation}
    \sigma_{TD} = \sigma_{Star} \sqrt{1 + \frac{1}{n_{_{T_{14}}}}}
\end{equation}
where n$_{\rm T_{\rm 14}}$ is the number of transit durations observed out of transit (i.e. the baseline). For all simulations presented here, $n_{T_{14}}$ is set to 2 (i.e. 1 x T$_{14}$ is spent both before/after the main observation). The error is calculated in this way for every spectral bin.

\subsubsection{Zodiacal Emission}
\label{sec:zodi}

We calculate the contribution of zodiacal emission using the prescription from \cite{pascale_echosim} and \cite{sarkar_jexosim}. The signal is composed of two black bodies, with associated coefficients, to model the reflected and emitted components. The spectral brightness is given by:

\begin{equation}
\begin{aligned}
Zodi(\lambda) = \beta (3.5\times10^{-14}B_\lambda (5500 K) \\
+ 3.58\times10^{-8}B_\lambda (270 K) )
\end{aligned}
\end{equation}
where the coefficient $\beta$ modifies the intensity of the zodiacal light based upon the declination of the target. At the ecliptic poles, $\beta$ = 1 provides a good fit to the intensity shown in \cite{leinert_zodi}. \cite{sarkar_jexosim} fitted a polynomial to data from this study, along with zodiacal intensities from \cite{james_zodi,tsumura_zodi}, to provide a measure of the increase in intensity at different latitudes. If d is the ecliptic latitude, then the coefficient is given by:
\begin{equation}
\begin{aligned}
    \beta = -0.22968868\zeta^7 + 1.12162927\zeta^6 - 1.72338015\zeta^5\\
    + 1.13119022\zeta^4 - 0.95684987\zeta^3 + 0.2199208\zeta^2\\
    - 0.05989941\zeta + 2.57035947
\end{aligned}
\end{equation}
where $\zeta$ = $log_{10}(d + 1)$. This relation falls below 1 at d = 57.355 $^\circ$ and so $\beta$ is fixed to 1 for latitudes greater than this \citep{sarkar_jexosim}.

\subsubsection{Pointing Jitter}
\label{sec:jitter}

Directly modelling uncertainties due to spacecraft jitter is beyond the capabilities of Terminus. Hence, ExoSim has been adapted to the Twinkle design to study the effects of pointing jitter on science performance. ExoSim, first conceived for EChO \cite{pascale_echosim} and now used for the Ariel mission, has previously been adapted for simulating observations with JWST \citep{sarkar_jexosim} and the EXoplanet Climate Infrared TElescope (EXCITE, \cite{tucker_excite,nagler_excite}). The modified version, christened TwinkleSim, was run for a number of stellar types (T$_{\rm S}$ = 3000, 5000, 6100 K) and magnitudes (K$_{\rm S}$ = 6, 9, 12) and the uncertainty due to jitter determined in each case. Twinkle's baseline pointing solution is based upon a high performance gyroscope and a Power Spectral Density (PSD) was supplied by the engineering team at the satellite manufacturer, Airbus. For each simulation, a variety of different extraction apertures were trialled with larger apertures reducing the jitter by ensuring clipping did not occur but increasing the noise from other sources due to sampling more pixels (e.g. dark current). After trialling a number of solutions, the aperture was set to be rectangular with a width of 2.44 times the width of the Airy disk at longest wavelength of each channel. In terms of pixels, this is equivalent to 12 and 22 in the spatial direction, for Ch0 and Ch1 respectively, while the spectral pixels per bin are set to 6 and 7.

When combining observations time-correlated noise may integrate down more slowly than uncorrelated noise, which is assumed to decrease with the square root of the number of observations, and thus can contribute more heavily to the final noise budget. To account for this Allan deviations plots were produced using TwinkleSim. A power law trend can be fitted to this and used to derive a wavelength-dependent fractional noise term that jitter induces on the photon noise. For more details on this process, we refer the reader to \cite{sarkar_jexosim}.


\subsubsection{Transit Signal}
During transit, the critical signal is the fraction of stellar light that passes through the atmosphere of the exoplanet. This signal is determined by the ratio of the projected area of the atmosphere to that of the stellar disk and thus is given by:

\begin{equation}
\frac{2R_p \Delta z(\lambda)}{R_*^2}
\end{equation}
where $\Delta z$ is the height of the atmosphere. The size of the atmosphere is taken to be equivalent to the height above the the 10 bar radius, at which point the atmosphere is assumed to be opaque. The pressure of an atmosphere at a height, z, is given by:
\begin{equation}
p(z) = p_0 e^{\frac{-z}{H}}
\end{equation}
where H is the scale height, the distance over which the pressure falls by 1/e. In the literature, 5 scale heights are often assumed for $\Delta z$ for a clear atmosphere (at which point one is above 99.5$\%$ of the atmosphere) while 3 would be more reasonable in the moderately cloudy case \citep{puig,tinetti_ariel,edwards_ariel}. The scale height of the atmosphere is calculated from:
\begin{equation}
H = \frac{kT_pN_A}{\mu g}
\end{equation}
where k is the Boltzmann constant, $N_A$ is Avogadro's number, $\mu$ is the mean molecular weight of the atmosphere and g is the surface gravity determined from:
\begin{equation}
g = \frac{GM_p}{R_p^2}
\end{equation}
where $M_p$ and $R_p$ are the mass and radius of the planet and G is the gravitational constant.

\subsubsection{Eclipse Signal}

During eclipse, the signal is calculated form two sources; reflected and emitted light from the planet. Emission from the exoplanet day-side is modelled as a black body and the wavelength-dependent surface flux density is given by:
\begin{equation}
S_p(\lambda,T_p) = \pi \frac{2hc^2}{\lambda^5} \frac{1}{e^{\frac{hc}{\lambda kT_p}} -1}
\end{equation}
where $T_p$ is the dayside temperature of the planet. The product of the black body emission and the solid angle subtended by the exoplanet at the telescope gives the spectral radiance at the aperture: 
\begin{equation}
E_p^{Emission}(\lambda, T_p) = S_p(\lambda, T_p) \left( \frac{R_p}{d} \right)^2
\end{equation}
in $Wm^{-2}\mu m^{-1}$. Additionally, a portion of the stellar light incident on the exoplanet is reflected. The strength of this reflected signal is strongly dependant on wavelength and can be significant at visible wavelengths. The flux of reflected light at the telescope aperture is calculated from:
\begin{equation}
E_p^{Reflection}(\lambda) = \alpha_{geom} S_s(\lambda) \left( \frac{R_*}{d} \right)^2 \left( \frac{R_p}{a} \right)^2 
\end{equation}
where $S_S(\lambda)$ is the star spectral irradiance, $a$ is the star-planet distance (i.e. the planet's semi-major axis) and $\alpha_{geom}$ is the geometric albedo, which is assumed to be that of a Lambertian sphere ($\frac{2}{3} \alpha_{bond}$), wavelength-independent and at a phase of $\phi$ = 1 (i.e. full disk illumination). 

\subsubsection{Signal-to-Noise Ratio}

From these equations, and the error on the transit/eclipse depth, the signal-to-noise (SNR) on the atmospheric signal can be obtained for a single observation. Assuming the SNR increase with the square root of the number of observations, the SNR after multiple transits/eclipses is given by:
\begin{equation}
    SNR_N = \sqrt{N} SNR_1
\end{equation}
where SNR$_1$ is the SNR of a single observation and N is the total number of observations. By setting a requirement on the SNR (SNR$_R$), the number of observations needed for a given planet can be ascertained from:
\begin{equation}
    N = \left( \frac{SNR_R}{SNR_1} \right)^2
    \label{eq:num_obs}
\end{equation}
The current requirements are set to a median SNR $>$ 7 across 1.0-4.5 $\mu$m for transit observations and 1.5-4.5 $\mu$m for eclipse measurements. In the former of these the shorter wavelengths are excluded to avoid biasing against planets around cooler stars while the latter is chosen as planetary emission, even for relatively hot planets ($\sim$1500 K), is low at wavelengths shorter than 1.5 $\mu$m. Using Equation \ref{eq:num_obs}, one can then determine the type(s) of observation the planet is suited to.


\subsection{Atmospheric Modelling}
\label{atmo_modelling}

To simulate transmission (and emission) forward models, the open-source exoplanet atmospheric retrieval framework TauREx 3 \citep{al-refaie_taurex3, waldmann_2,waldmann_1} is used. Within TauREx 3, cross-section opacities are calculated from the ExoMol database \citep{yurchenko} where available and from HITEMP \citep{rothman} and HITRAN \citep{gordon} otherwise. The H- ion is included using the procedure outlined in \cite{john_1988_h-,ares1}. For atmospheric chemistry, two options are available within the Terminus infrastructure: chemical equilibrium, which is achieved using the ACE code \citep{venot_ace,agundez_ace} and takes the C/O ratio and metallicity as input, or free-chemistry which allows the user to choose molecules and their abundances. Alternatively, a high-resolution spectrum produced by another radiative transfer code can be read in or, if a retrieval on actual data has been performed, the atmosphere can be extrapolated from a TauREx 3 hdf5 file. Once the forward model is created at high resolution, it is then binned to the instrument resolution using TauREx 3's integrated binning function.

\subsection{Light Curve Modelling and Fitting}
\label{lc_modelling}

For each spectral bin, PyLightCurve\footnote{\url{https://github.com/ucl-exoplanets/pylightcurve}} \citep{tsiaras_plc} is used to model a noise-free transit/eclipse of the planet. The transits were all modelled with quadratic limb darkening coefficients from \cite{claretII}, calculated using ExoTETHyS \citep{morello_exotethys}. The Twinkle spectrometer features a split at 2.43 $\mu$m, creating two channels. For each of these a white light curve is also generated. The spectral light curves are created at the native resolution of the instrument (R$\sim$20-70). A time-series is created with a cadence equal to the time between destructive reads and the light curve integrated over each of these exposures. The noise per integration, as calculated in Section \ref{radiometric}, is then used to create noisy light curves by adding Gaussian scatter. Further updates will include the ability to add ramps due to detector persistence as well as other time-varying systematics.

For the fitting of the light curves a Markov chain Monte Carlo (MCMC) is run using emcee \citep{emcee} via the PyLightCurve package, here with 150,000 iterations, a burn-in of 100,000, and 100 walkers. For the simulations shown here, both white light curves are individually fitted with the inclination (i), reduced semi-major axis (a/R$_*$), transit mid-time (T$_0$) and planet-to-star radius ratio (R$_p$/R$_s$) as free parameters. A weighted average of the recovered values for each of these parameters, except the planet-to-star radius ratio, is then fixed for the fitting of the spectral light curves where only the planet-to-star radius ratio is fitted. This provides a retrieved transit/eclipse depth for each light curve, along with the error associated with this parameter. If further complexity, such as ramps, is added to the light curve, future iterations of the code will allow for multiple light curve fits. In this case the uncertainties in the individual data points are increased such that their median matches the standard deviation of the residuals, a common technique when analysing Hubble observations of exoplanets \citep[e.g.][]{kreidberg_gj1214,tsiaras_hd209}.

For fainter targets, a spectrum with a reduced resolution can be requested and Terminus will combine the light curves and provide a spectrum with a resolution as close to the desired as possible. While the default cadence is set by the saturation time of the detector it can lowered or exposures can be combined. Additionally, multiple transits (or eclipses) can be individually modelled, fitted and then combined. These functionalities are all controlled by the input configuration file. Once a spectrum has been generated, an automated interface with TauREx 3 can then be used to fit the data and retrieve the atmospheric parameters.

To compare the errors predicted by the radiometric model to those from fitted light curves, we model a single observation of HD\,209458\,b \citep{charbonneau_hd209,henry_hd209}. For the atmosphere we model a composition based loosely on that retrieved from the HST data of this planet \citep{tsiaras_hd209,macdonald}. We assume a plane parallel atmosphere with 100 layers and include the contributions of collision-induced absorption (CIA) of H$_2$-H$_2$ \citep{abel_h2-h2,fletcher_h2-h2} and H$_2$-He  \citep{abel_h2-he}, Rayleigh scattering and grey-clouds. In terms of molecular line lists, we import the following: H$_2$O \citep{polyansky_h2o},  NH$_3$ \citep{ExoMol_NH3}, CH$_4$ \citep{ExoMol_CH4_new} and HCN \citep{ExoMol_HCN}. 

\begin{figure}
    \centering
     \includegraphics[width=\columnwidth]{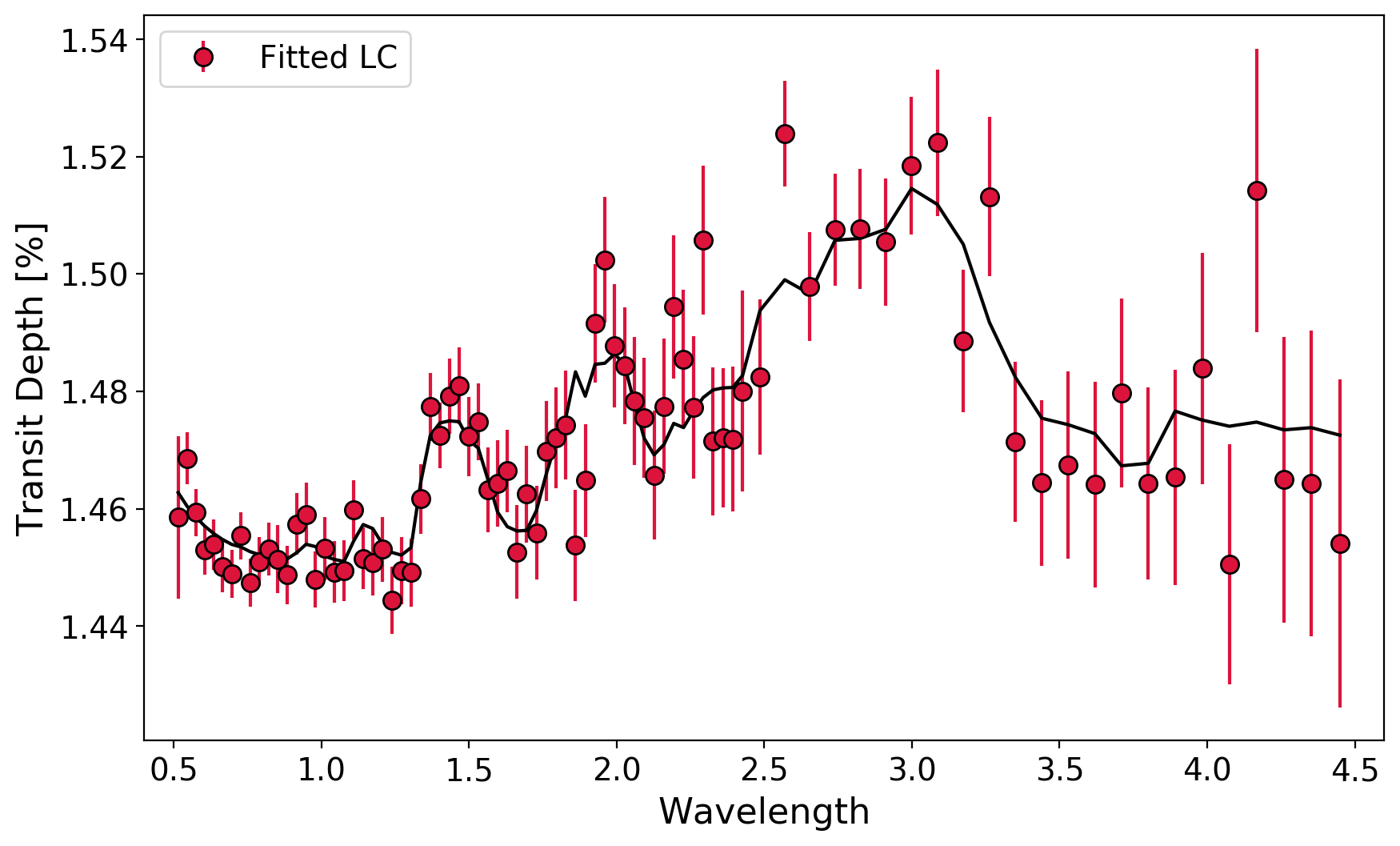}
    \includegraphics[width=\columnwidth]{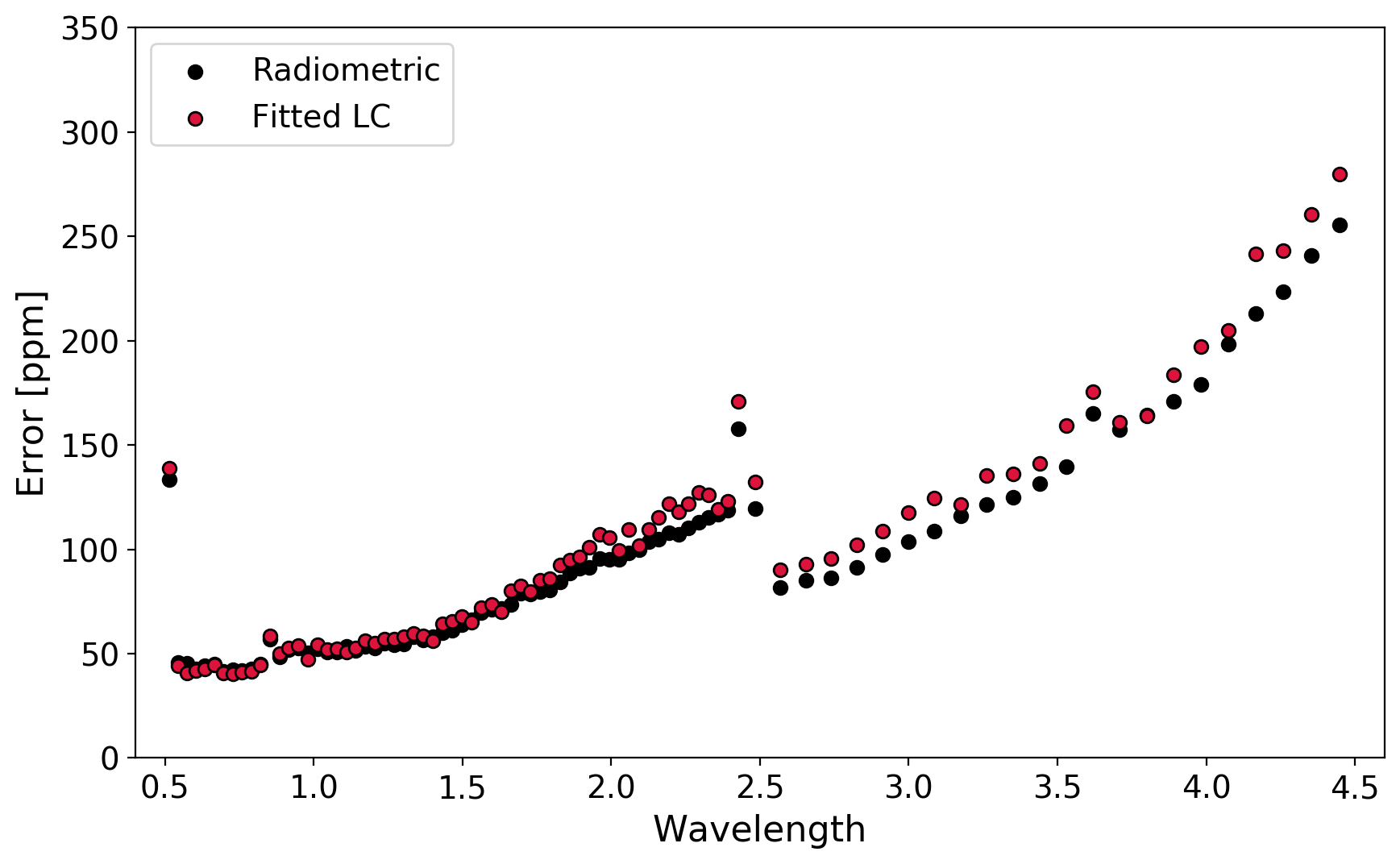}
    \includegraphics[width=\columnwidth]{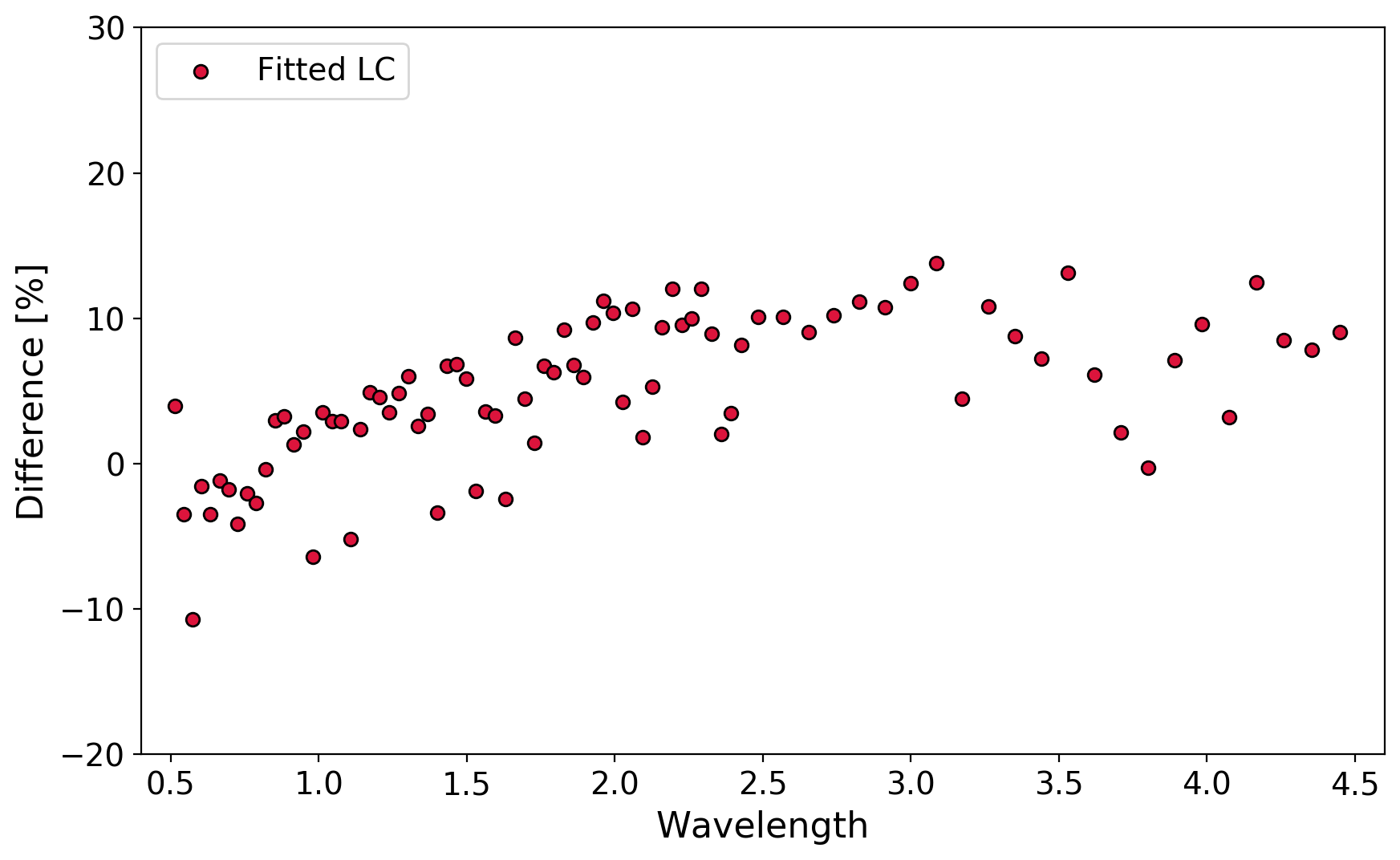}
    \caption{Comparison of error bars obtained from the radiometric model (black) and light curve fitting for HD\,209458\,b. The wavelength dependent difference between the models could be due to limb darkening coefficients.}
    \label{fig:hd209_lc_errors}
\end{figure}

Figure \ref{fig:hd209_lc_errors} displays the errors on the transit depth predicted by the radiometric portion of Terminus as well as the uncertainties recovered from the light curve fits. While the agreement is generally good, within 10\%, there appears to be a wavelength-dependent effect on the accuracy of the radiometric tool. The trend seen could be due to the limb darkening coefficients, which change with wavelength and alter the shape of the light curve.

\section{Orbit Modelling}
\label{obstruction}

Observatories in low Earth orbits can experience interruptions in target visibility due to Earth occultations. Additionally, instruments and spacecraft usually have specific target-Sun, target-Moon or target-Earth limb restrictions. To account for these, Terminus is capable of modelling the orbit of a spacecraft and calculating angles between the target and the Earth limb, the Sun or other celestial body, in a similar way to tools used for other missions (e.g. for CHEOPS: \cite{kuntzer_salsa}). 

The tool operates within an Earth-centred frame and the positions of celestial objects (the Sun, Moon etc.) are loaded from the JPL Horizons service\footnote{\url{https://ssd.jpl.nasa.gov/horizons.cgi}}. The spacecraft's orbit is defined by an ellipse which is subsequently inclined with respect to the X plane. The right ascension of the ascending node (RAAN) is then used to rotate this about the Z axis.

\begin{figure}
    \centering
    \includegraphics[width = \columnwidth]{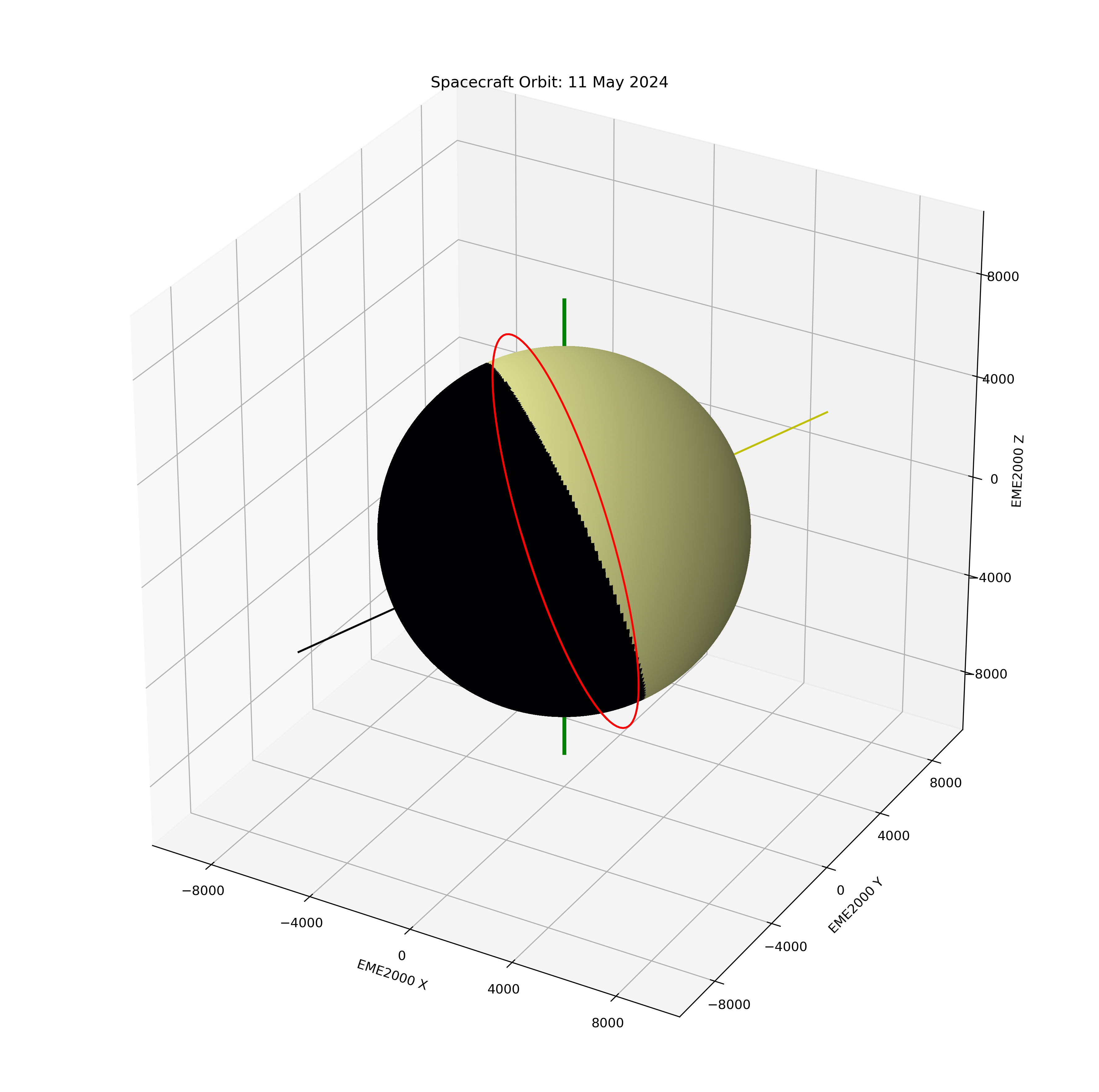}
    \caption{Modelled orbit of Twinkle (red) during May 2024. The yellow vector indicates the direction of the Sun while the black represents the anti-sun vector (i.e. the centre of Twinkle's field of regard). The Earth is represented by the sphere with the terminator between day and night roughly shown.}
    \label{fig:twinkle_orbit}
\end{figure}

Twinkle will operate in a Sun-synchronous orbit and here we modelled the following orbital parameters: altitude = 700 km, inclination = 90.4$^\circ$, eccentricity = 0, RAAN = 190.4$^\circ$ (i.e. 6am). These are subject to change based upon launch availability but provide an approximate description of the expected operational state. The orbit of Twinkle during May 2024 is depicted in Figure \ref{fig:twinkle_orbit}.

\begin{figure*}
    \centering
    \includegraphics[width=0.475\textwidth]{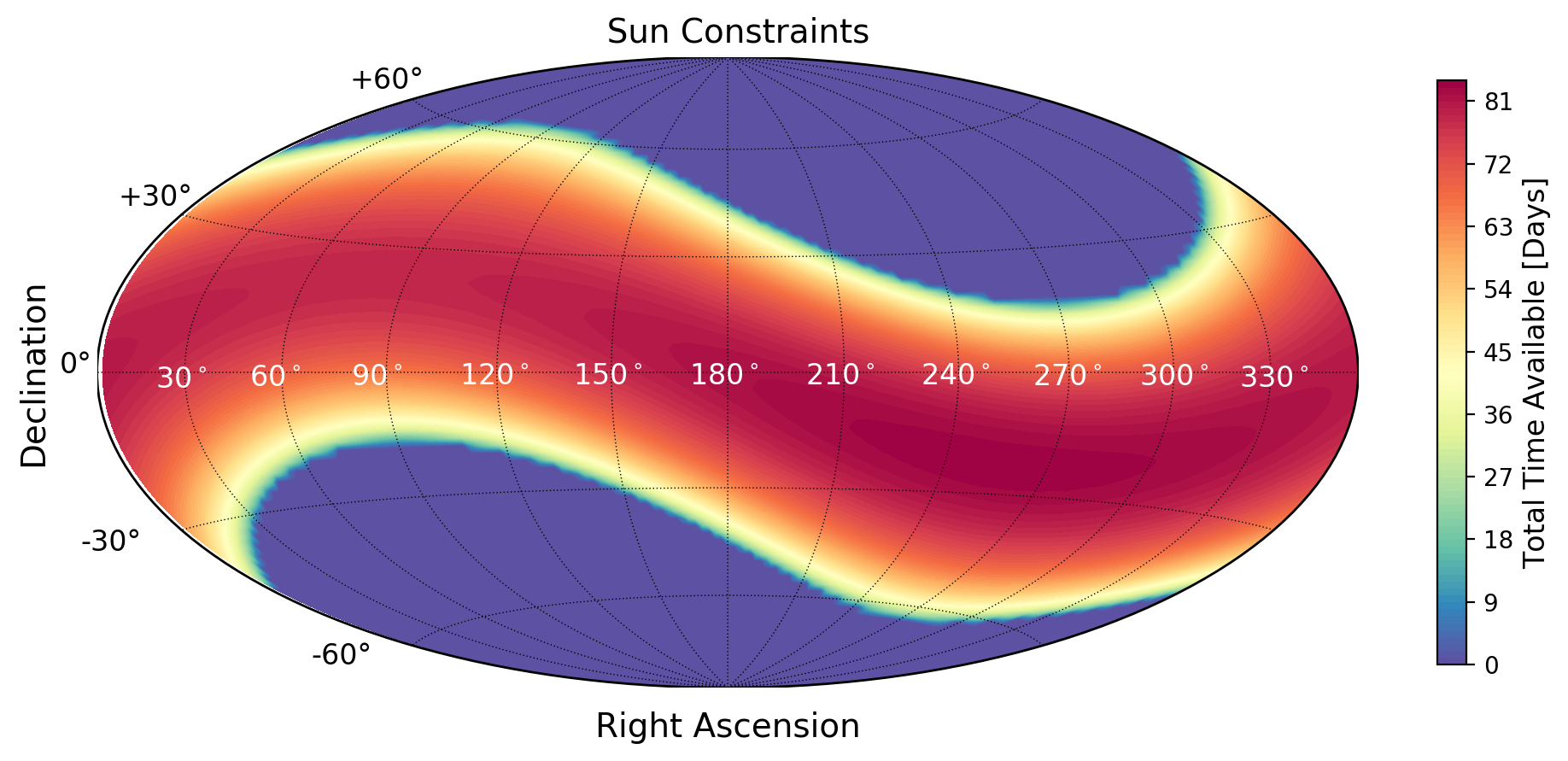}
    \includegraphics[width=0.475\textwidth]{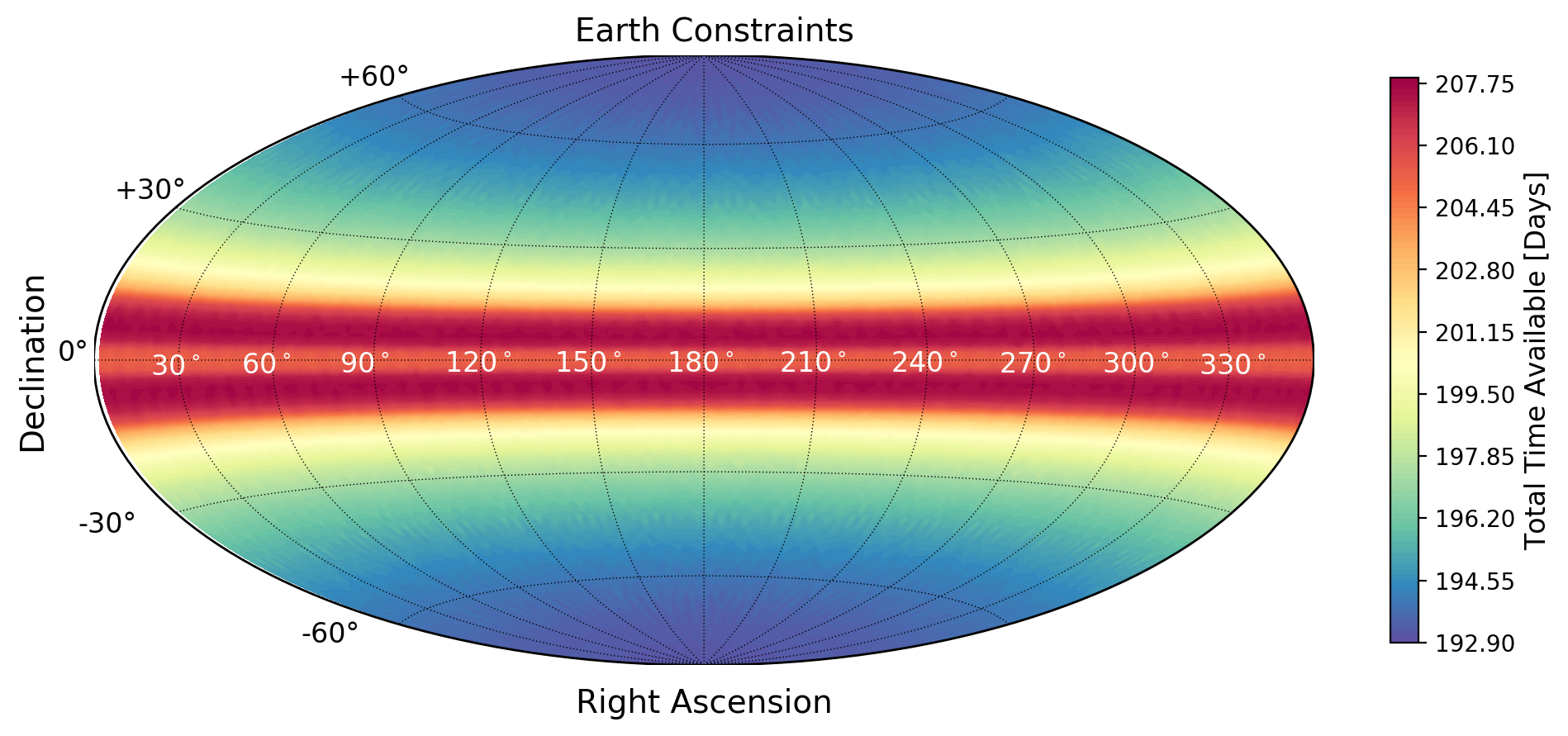}
    \includegraphics[width=0.475\textwidth]{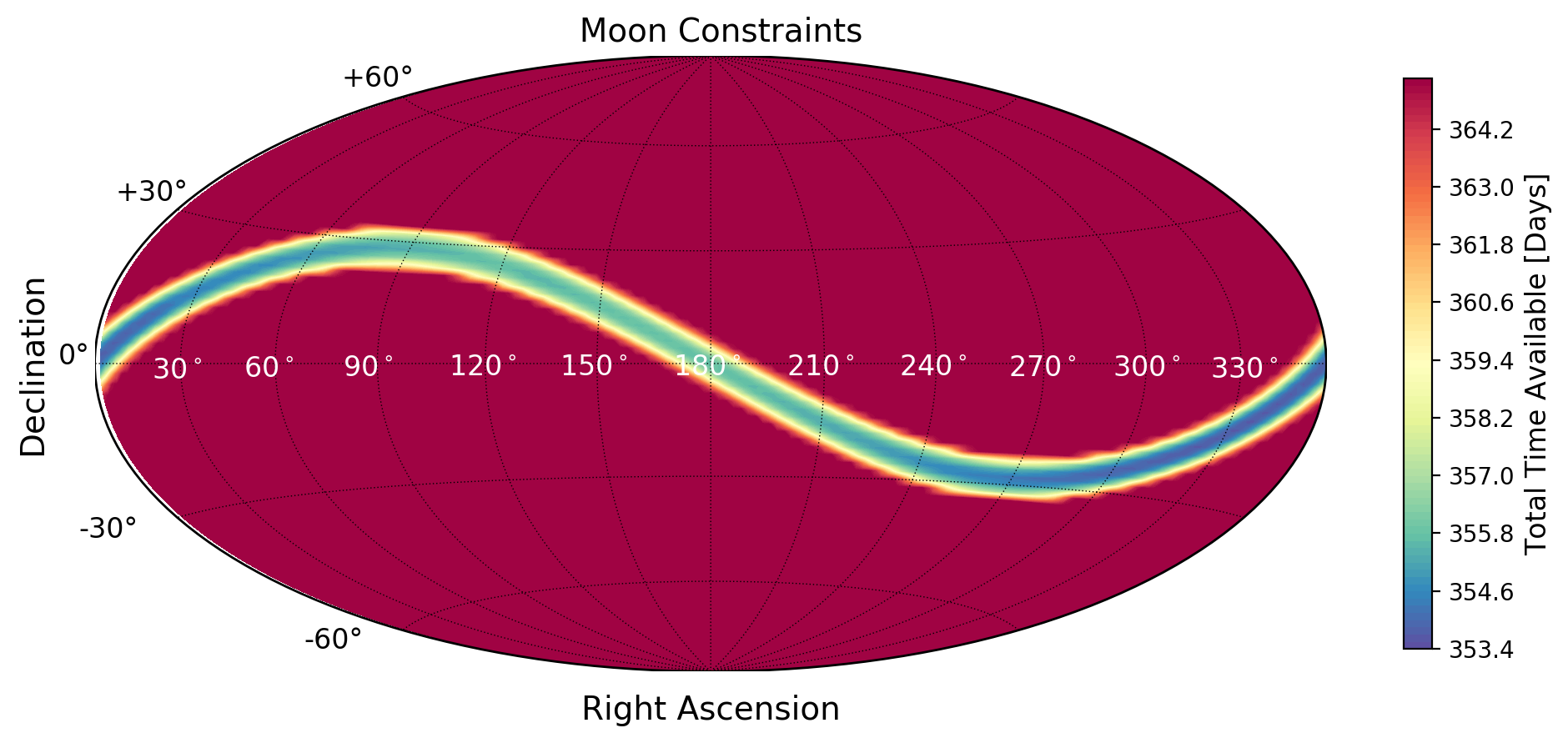}
    \includegraphics[width=0.475\textwidth]{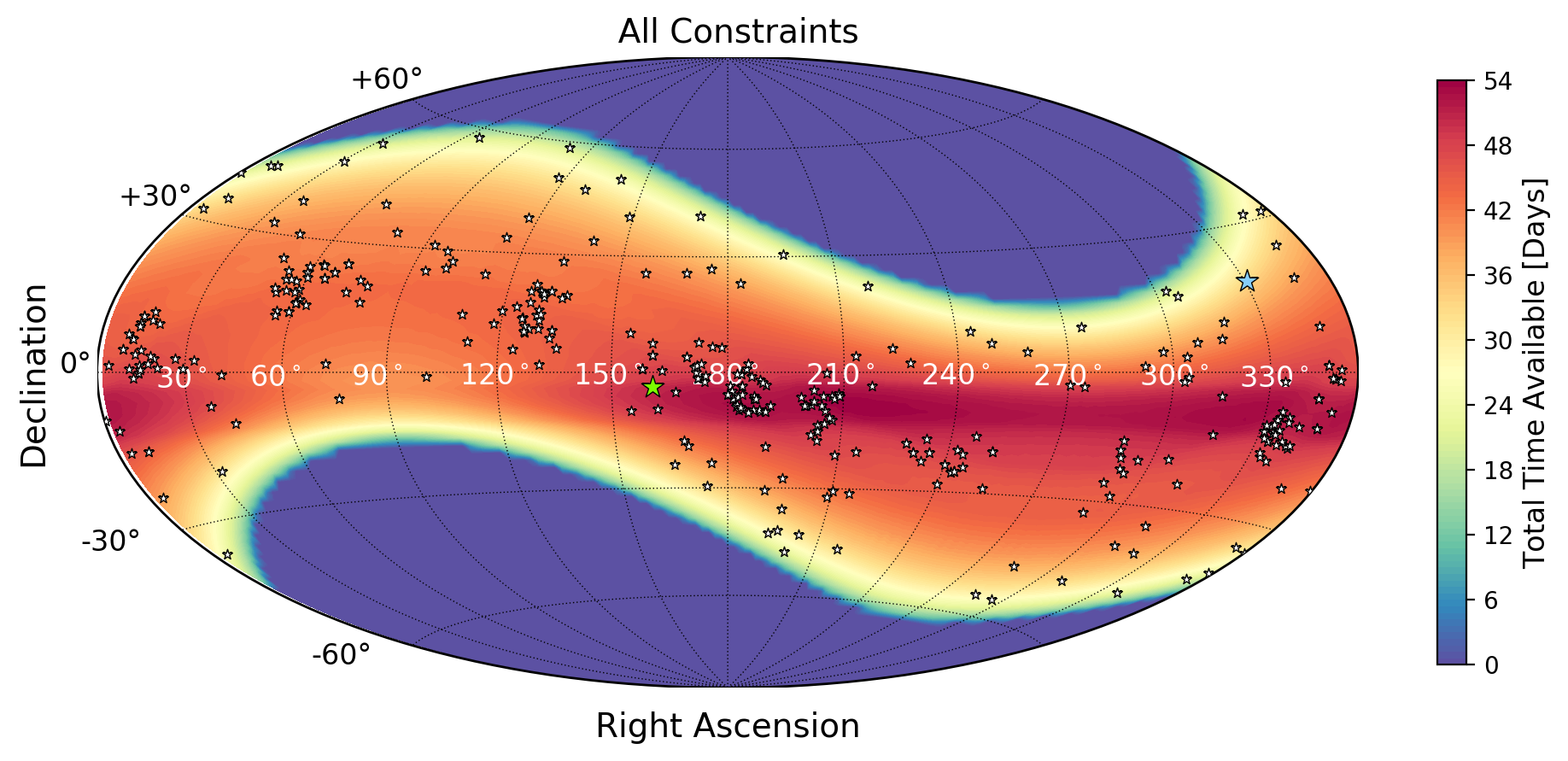}
    \caption{Sky coverage of Twinkle given the specific exclusion angles. The effects of individual constrains are shown for the Sun, Earth and Moon alongside the combination of them all. Stars indicate known transiting exoplanet hosts with HD\,209458 and WASP-127 highlighted by light blue and green stars respectively. We note that the colour bar axes differ between each plot.}
    \label{fig:sky_vis}
\end{figure*}

\begin{figure*}
    \centering
    \includegraphics[width=0.45\textwidth]{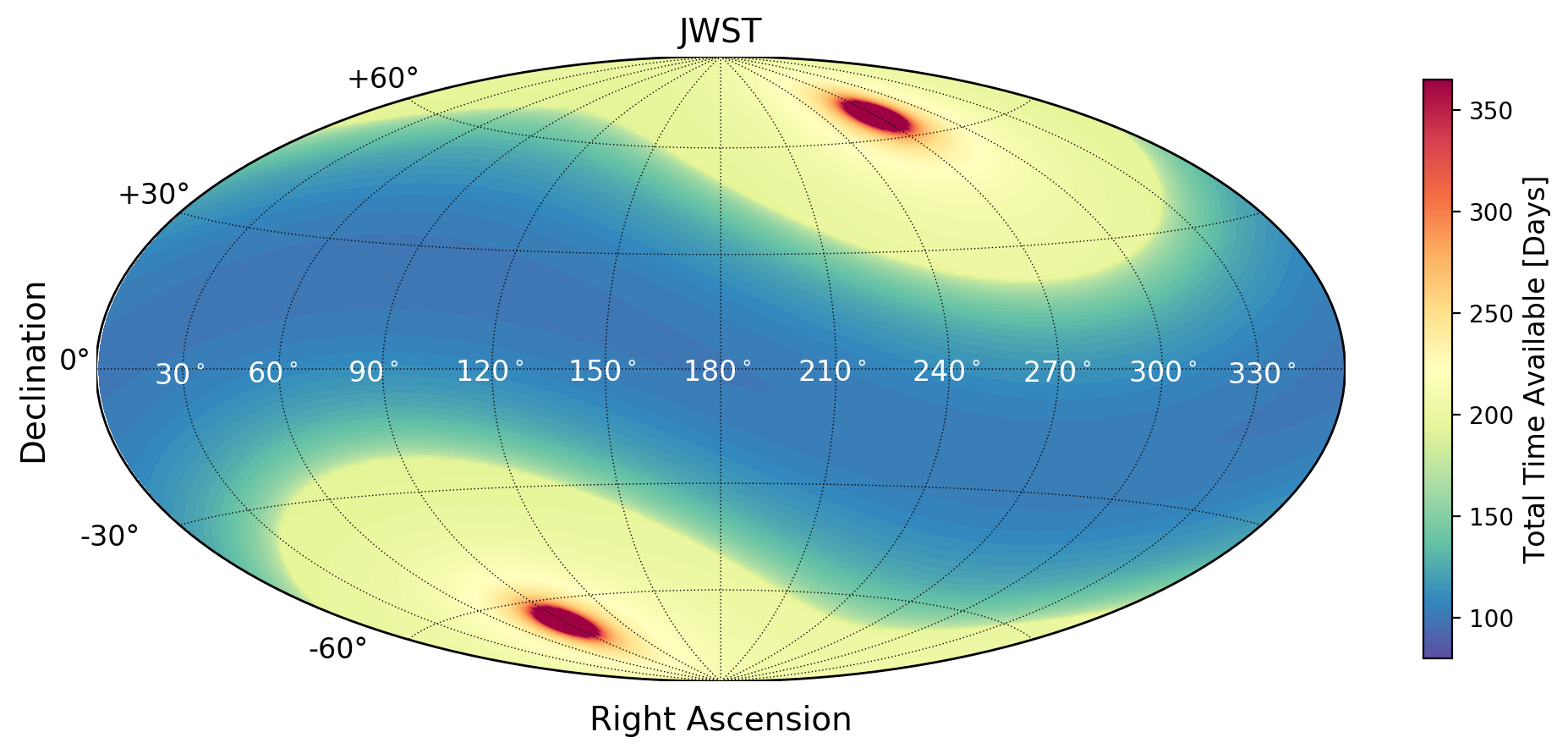}
    \includegraphics[width=0.45\textwidth]{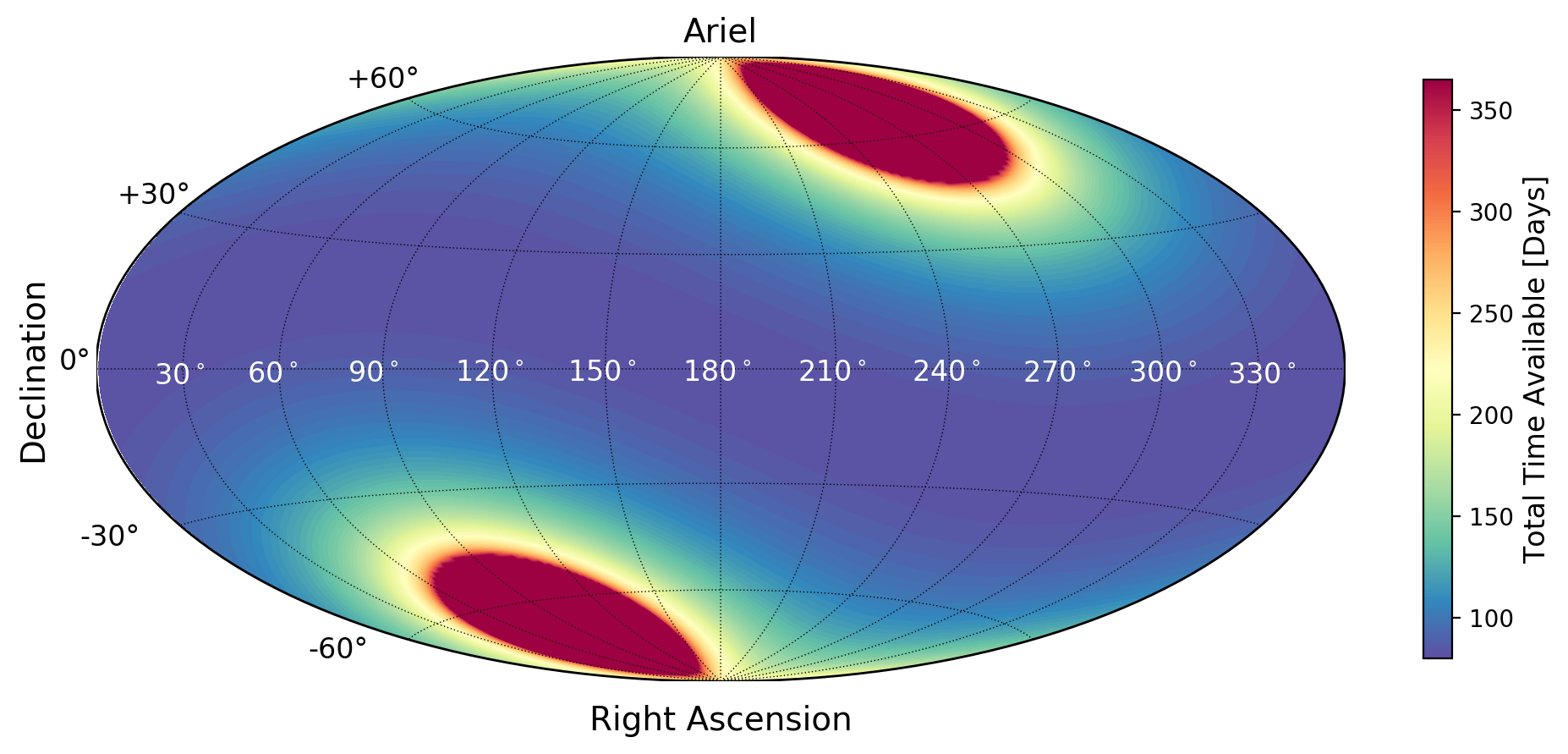}
    \caption{Sky coverage of JWST (left) and Ariel (right) which will have continuous viewing zones at the ecliptic poles. These missions are unaffected by Earth obscuration due to their L2 orbit.}
    \label{fig:jwst_ariel}
\end{figure*}

As mentioned, the code can impose a number of exclusion angles to explore their effects on target availability. Here we modelled Sun, Earth and Moon exclusion angles of 140 $^\circ$, 20 $^\circ$ and 5 $^\circ$ respectively. The first of these is largely due to thermal constraints while the latter two are to reduce stray light. The Earth and Moon exclusion angles for Twinkle are still under study but the values chosen here are similar to those of other observatories operating in sun-synchronous orbits or those proposed to do so \citep{kuntzer_salsa, swain_finesse}.

The effects of each exclusion angle on the sky coverage is shown in Figure \ref{fig:sky_vis} along with the effect of combining them all. In each case, the metric shown is the total time the area of sky can be observed over the course of a year. The plots highlight Twinkle's excellent coverage of the ecliptic plane although it, like CHEOPS, lacks the ability to study planets close to the ecliptic poles. However, the JWST and Ariel missions will prefer the polar regions, as shown in Figure \ref{fig:jwst_ariel}, and thus both Twinkle and CHEOPS provide complimentary coverage. 


\section{Partial Light Curves}

\begin{figure}
    \centering
    \includegraphics[width = \columnwidth]{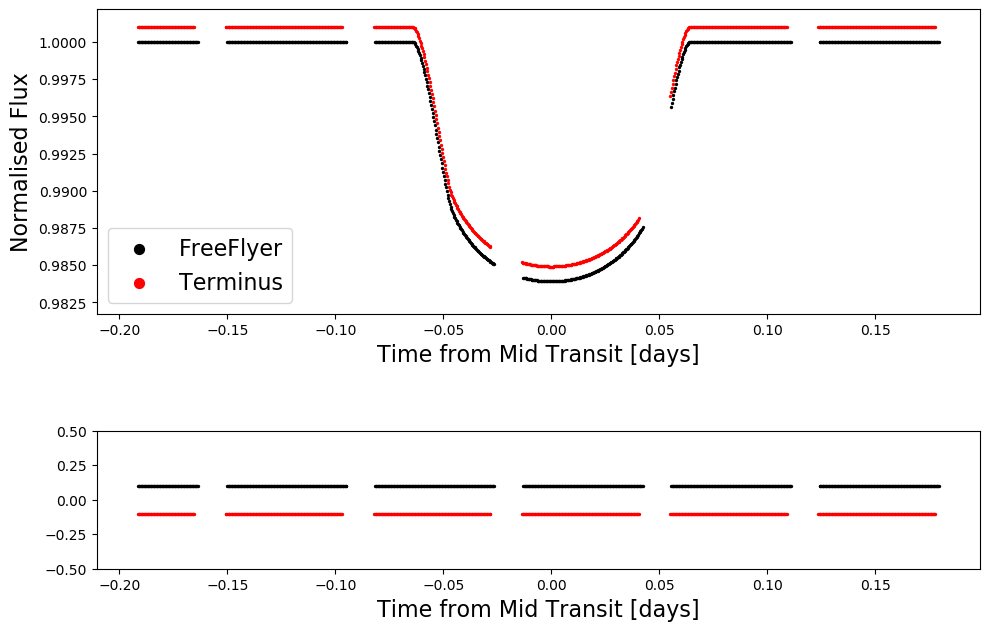}
    \caption{Comparison of the predicted gap sizes for HD 209458 b (RA = 330, Dec = 18) from Terminus and Freeflyer. The transit light curves are offset for clarity and the gap sizes are seen to be highly similar. We note that these gaps are due solely to physical obscuration by the Earth and no exclusion angle is included.}
    \label{fig:TF_comp}
\end{figure}

\begin{figure}
    \centering
    \includegraphics[width = \columnwidth]{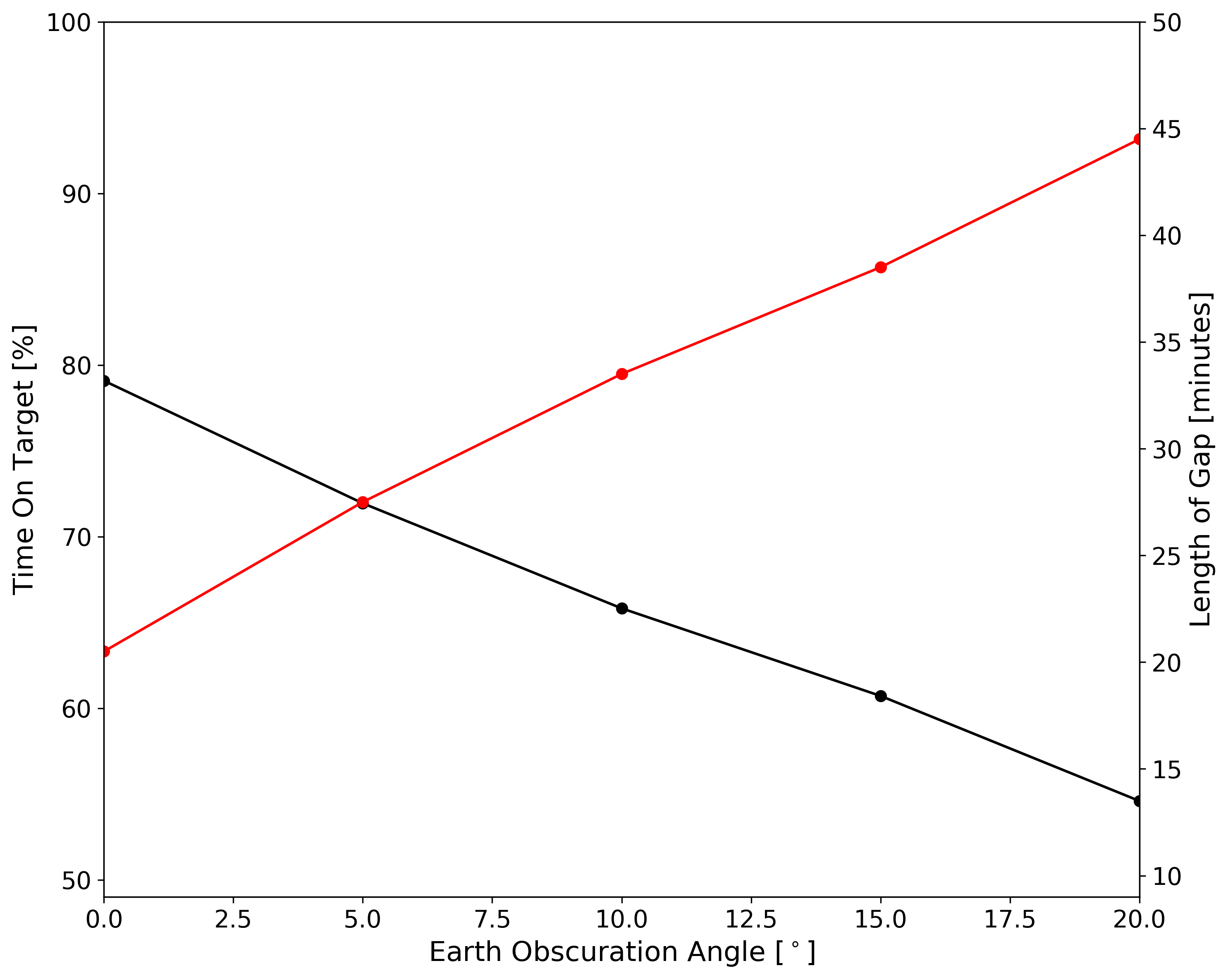}
    \caption{Effect of different Earth exclusion angles on the percentage of time on target (black) and size of the gaps (red) for a transit observation of HD 209458 b.}
    \label{fig:HD209_gaps}
\end{figure}

\begin{figure}
    \centering
    \includegraphics[width=\columnwidth]{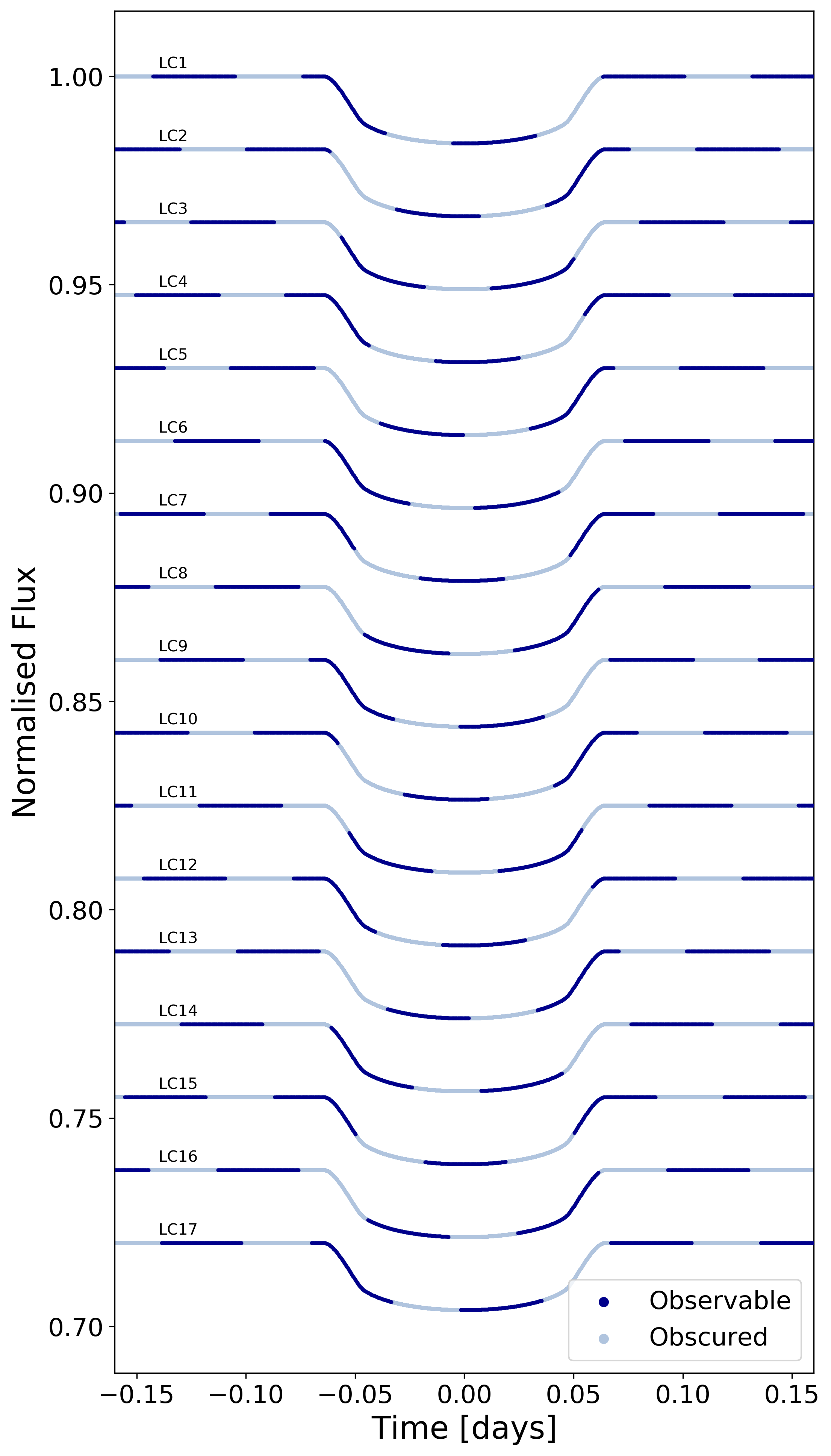}
    \caption{The 17 transits of HD\,209458\,b that are observable with Twinkle over the course of a single year. The gaps are due to Earth obscuration plus an exclusion angle of 20$^\circ$. All light curves have gaps of roughly 45 minutes which are comparable to those in the Hubble data of the same planet and have been offset for clarity.}
    \label{fig:hd209_transits}
\end{figure}

\begin{figure}
    \centering
    \includegraphics[width=\columnwidth]{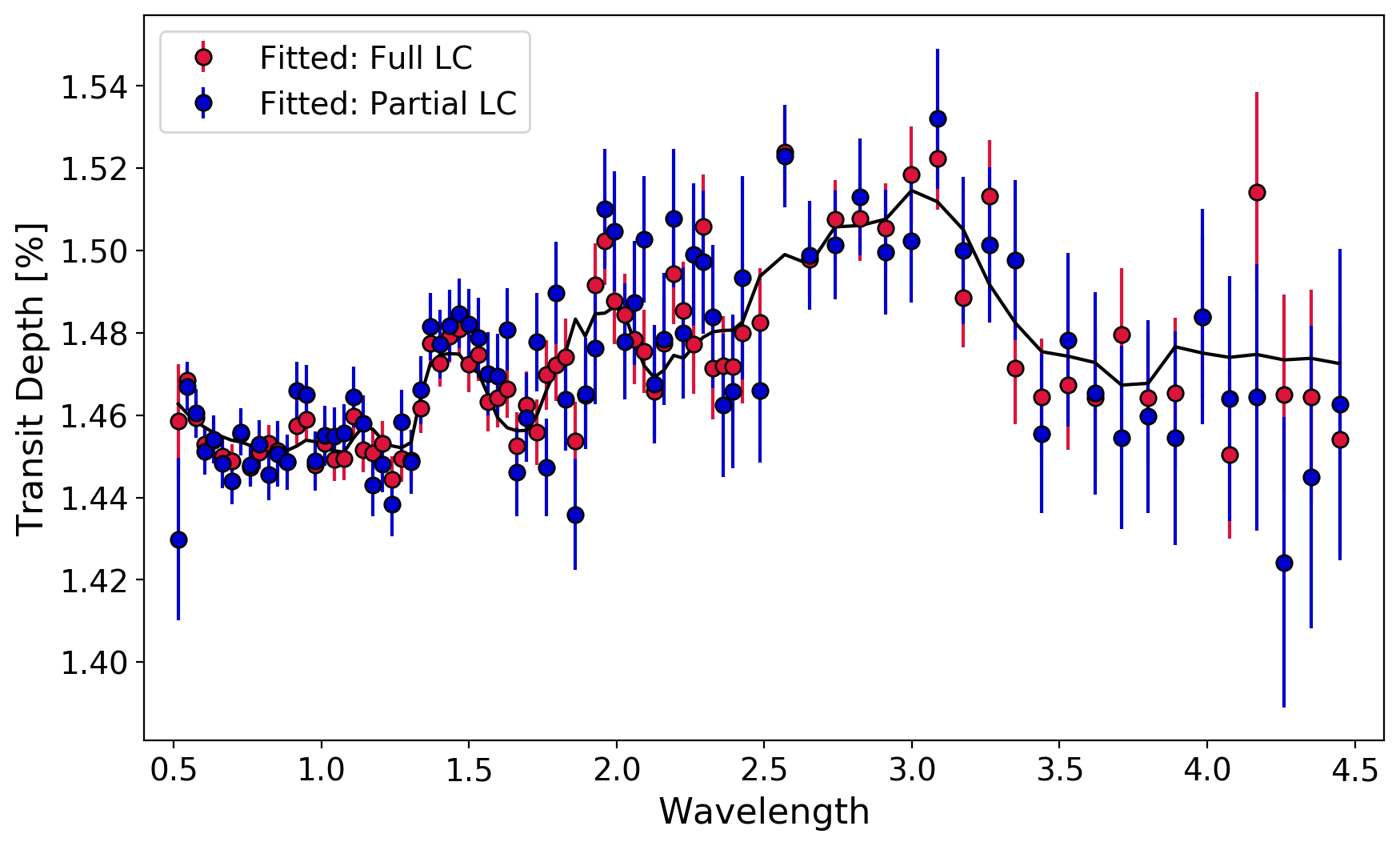}
    \includegraphics[width=\columnwidth]{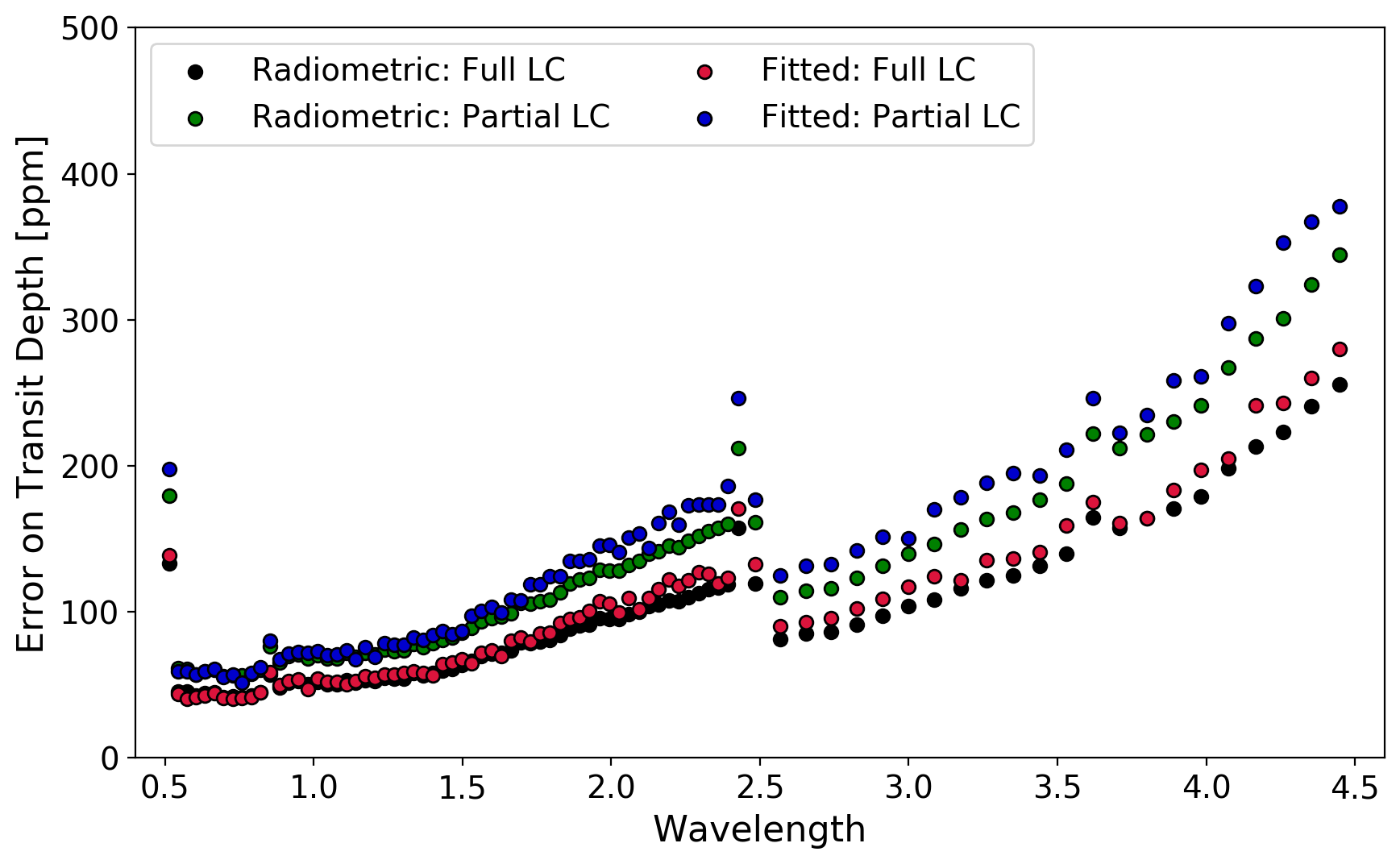}
    \includegraphics[width=\columnwidth]{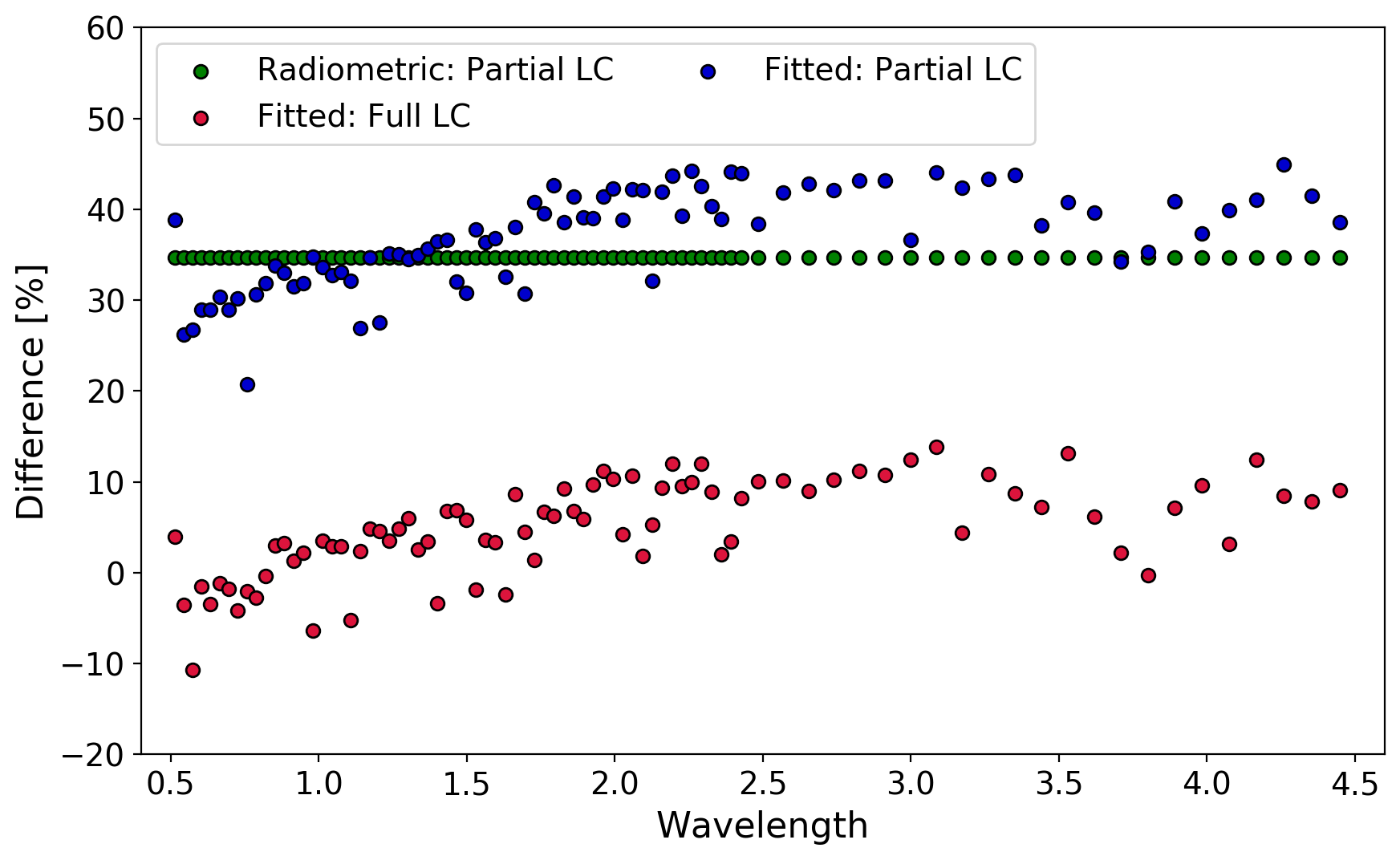}
    \caption{Recovered spectrum and error bars from different light curve fits for HD 209458 b. In each case, red represents the fitting of a full light curve (same as Figure \ref{fig:hd209_lc_errors}), blue the fitting of the partial light curve (LC1 from Figure \ref{fig:hd209_transits}) and black represents the predicted error from the radiometric model. The partial light curve results in far larger uncertainties due to the reduction in the number of data points.}
    \label{fig:hd209_errors}
\end{figure}

From an exoplanet modelling perspective, it has thus far it has been assumed that a full light curve is observed. However, in reality, for space-telescopes in a low-Earth orbit, sometimes only partial light curves will be obtained due to Earth obscuration as discussed in Section \ref{obstruction}. These gaps cannot be completely accounted for in radiometric models and thus a time-domain code, such as Terminus, is required. 

To verify the orbital code created, and to explore the effect of partial light curves, we check our results against those of \cite{billy_thesis}. In \cite{billy_thesis}, the mission design, analysis and operation software Freeflyer\footnote{\url{https://ai-solutions.com/freeflyer/}} was used to model the obscurations of HD\,209458\,b by the Earth throughout a year. FreeFlyer has previously been used to support planning for several missions including NASA’s Solar Dynamics Observatory (SDO). We note that Freeflyer only models the physical obscuration of the target star by Earth and thus for this comparison we set the Earth exclusion angle to zero.

As mentioned, Twinkle's field of regard means targets are not constantly observable and in a year 17 transits of HD\,209458\,b would be observable by Twinkle. Given the sky location of HD\,209458, Right Ascension (RA): 330.79$^\circ$; Declination (Dec): 18.88$^\circ$, the target will always be periodically obscured by the Earth. In Figure \ref{fig:TF_comp}, we show a comparison between the predicted gaps for the first of these transits which are shown to be in excellent agreement. Meanwhile, Figure \ref{fig:HD209_gaps} displays the increase in gap size that would be incurred by various Earth exclusion angles. Going from an angle of 0 to 20 degrees increases the gaps size from 20 minutes to 44 minutes. The latter case would mean Twinkle could be on-target for over half an orbit (54 minutes). In comparison, past Hubble observations featured gaps of 47 minutes, with 48 minutes on target per orbit \citep{deming_hd209,tsiaras_hd209}. Hence, Twinkle's observing efficiency for HD\,209458\,b will probably be similar to that of Hubble. All potential transit observations of HD\,209458\,b have gaps or a similar size (see Figure \ref{fig:hd209_transits}).

Here we fit the first available light curve and the recovered spectrum, and associated errors, is shown in Figure \ref{fig:hd209_errors}. As expected, the gaps increase the uncertainties on the recovered transit depth. Using Equations 5 and 8, one would expect the error to increase by 35\% ($\sigma_p = \sigma_f \times \frac{1}{\sqrt{0.55}} = 1.347 \sigma_f$). We see an increase of 20-40\% and thus the radiometric model may also provide reasonable errors for partial light curves.

\begin{figure}
    \centering
    \includegraphics[width=\columnwidth]{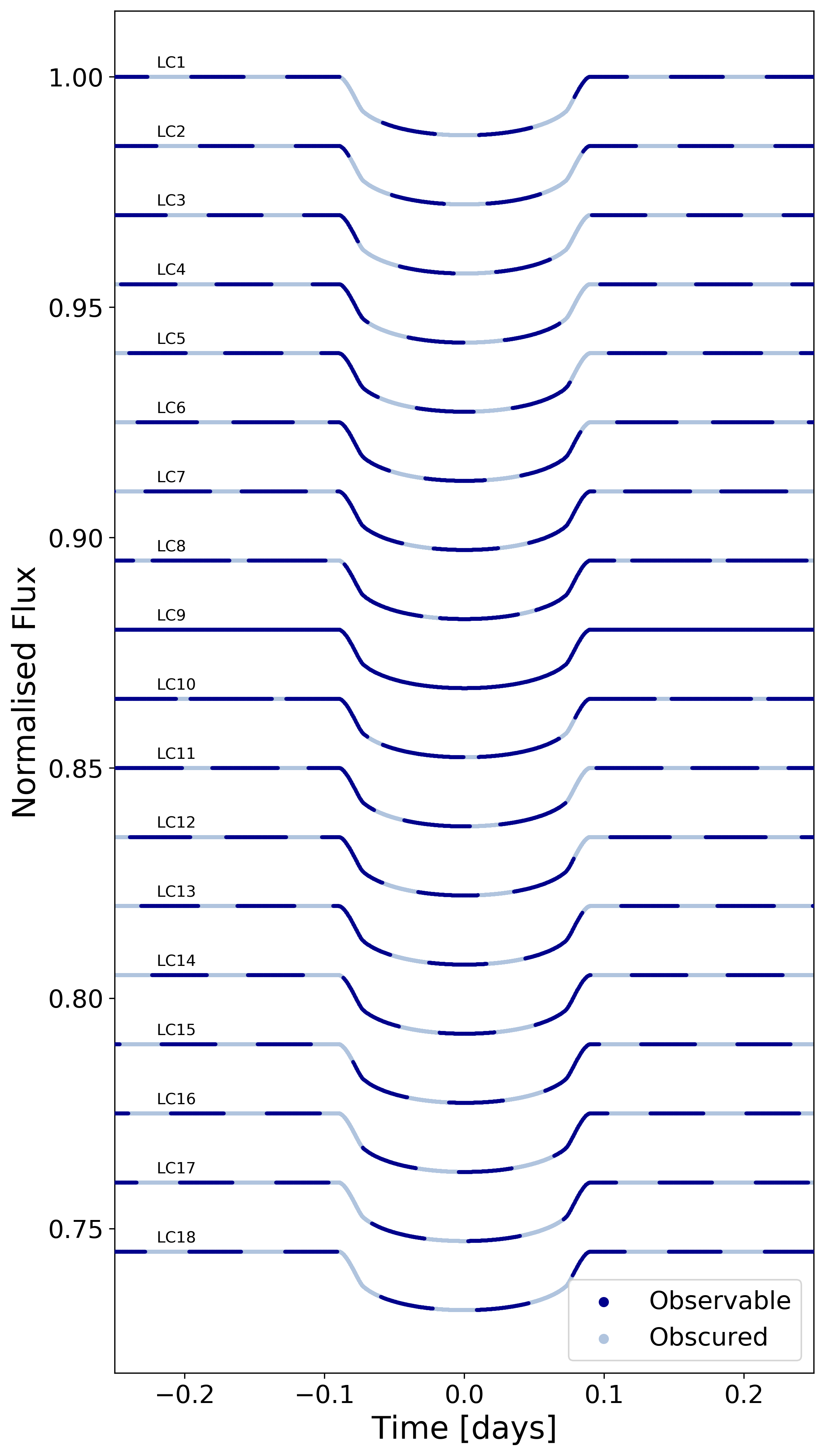}
    \caption{The 18 transits of WASP-127 b that are observable with Twinkle in 2024 which have been offset for clarity. The gaps are due to Earth obscuration plus an exclusion angle of 20$^\circ$. LC 9 has no gaps, highlighting the importance of observational planning with Twinkle, or other LEO satellites, and the benefit of a sun-synchronous orbit over the equatorial orbit of Hubble.}
    \label{fig:w127_transits}
\end{figure}

\begin{figure}
    \centering
    \includegraphics[width=\columnwidth]{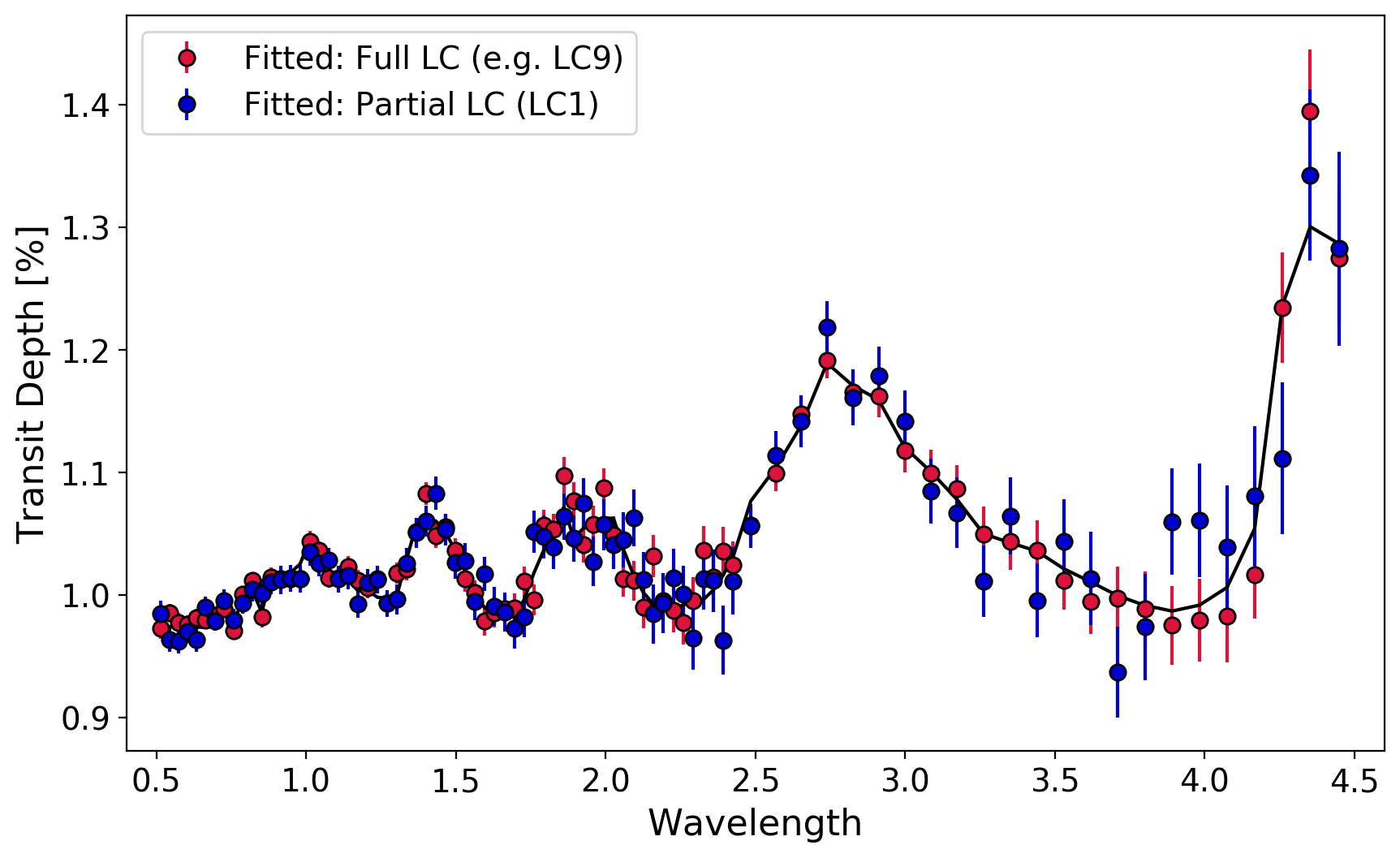}
    \includegraphics[width=\columnwidth]{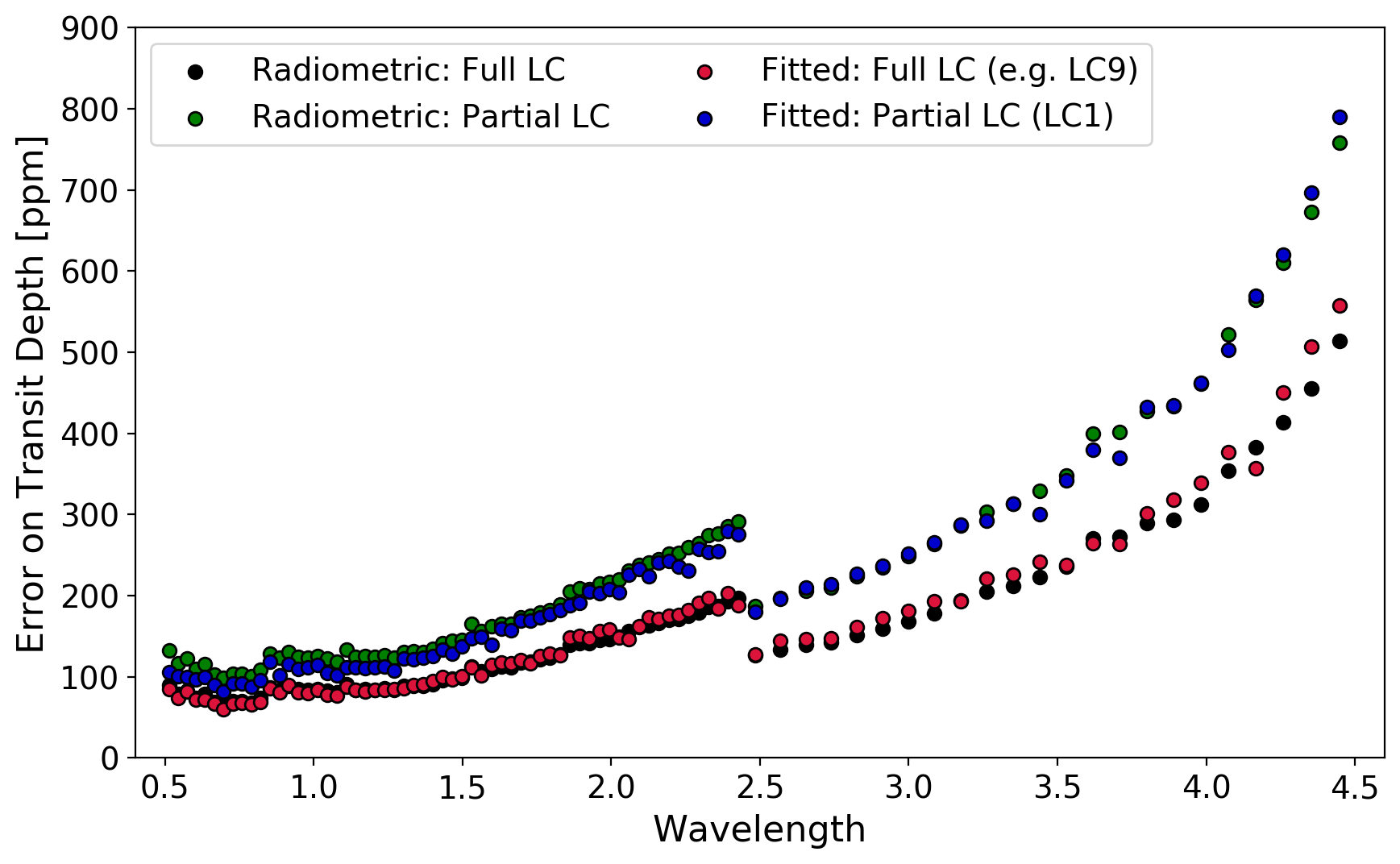}
    \includegraphics[width=\columnwidth]{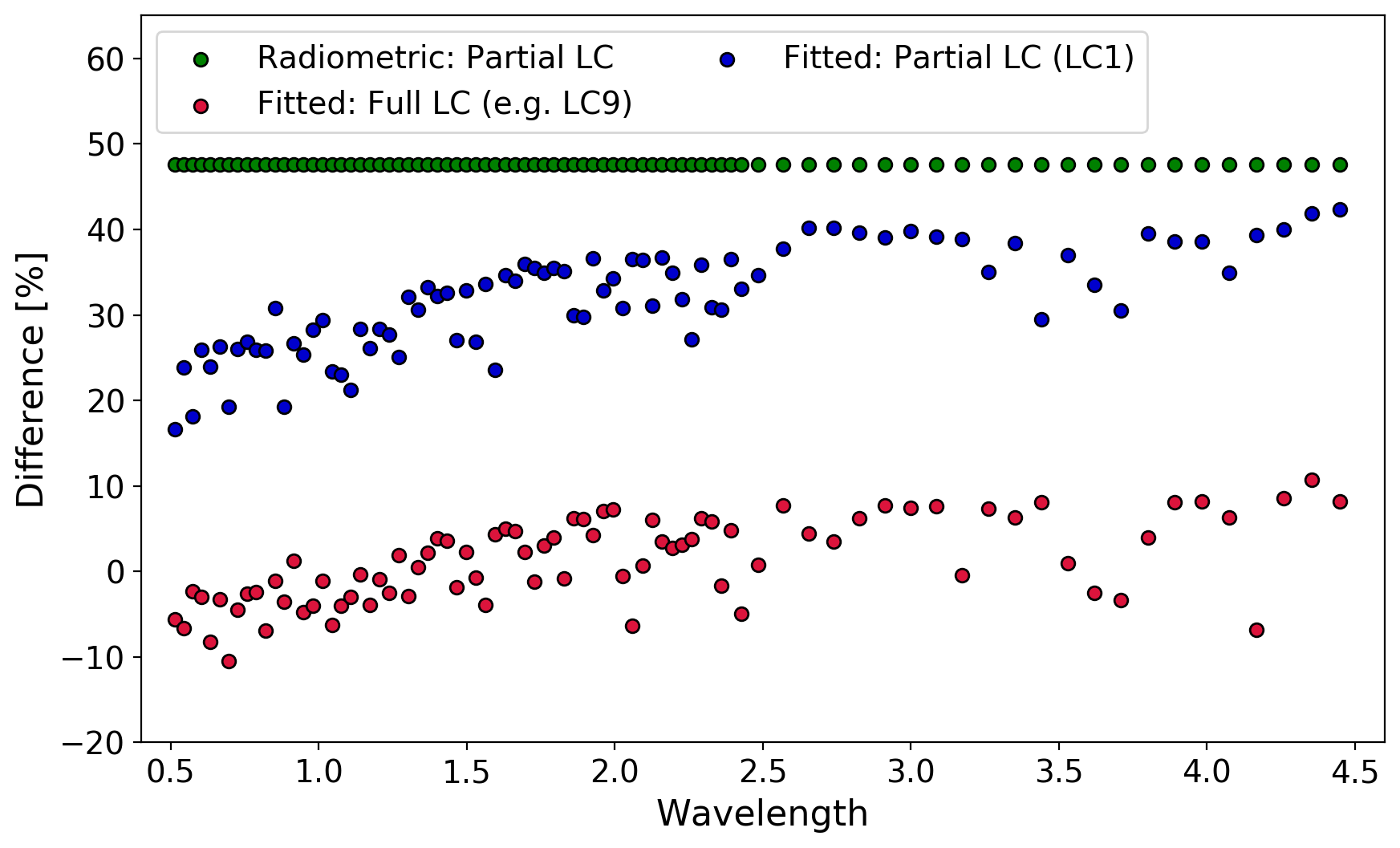}
    \caption{Recovered spectrum and error bars from different light curve fits for WASP-127\,b. In each case, red represents the fitting of a full light curve (e.g. LC9 in Figure \ref{fig:w127_transits}), blue the fitting of the partial light curve (LC1 from Figure \ref{fig:w127_transits}) and black represents the predicted error from the radiometric model. The errors from the full light curve are found to agree with the radiometric prediction, again with the exception of a slight, wavelength dependent, variation. The partial light curve results in far larger uncertainties.}
    \label{fig:wasp127_errors}
\end{figure}

However, some planets may have more variable gaps, due to their location in the sky and a changing spacecraft-Earth-target geometry, and thus may be affected more significantly. For these planets, the scheduling of observations is likely to be highly important. Terminus is able to provide input into studies exploring the effects of partial light curves.

As an initial step to understand the variability of Earth obscuration, we now model observations of WASP-127\,b \citep{lam_w127}. WASP-127 is located such that Twinkle will potentially have a continuous, unobstructed view of the target during a transit (RA: 160.56$^\circ$, Dec: -3.84$^\circ$). However, some potential observations will incur Earth obscuration and the amount of time lost will be dependent upon the Earth exclusion angle required. In the case of the 20$^\circ$ exclusion angle modelled here, Twinkle would have access to one complete transit (i.e. no gaps due to Earth obscuration) in 2024 as shown in Figure \ref{fig:w127_transits}. The other available observation periods would incur interruption up to a maximum of 45 minutes over a 98 minute orbit. In the case of the Hubble observations of WASP-127\,b \citep{skaf_w127,spake_w127}, the spacecraft could only be pointed at the target for 40 minutes per orbit (55 minute gaps). Hence, through careful selection of observing windows, the efficiency of Twinkle's observations of WASP-127\,b could be far greater than that of Hubble's for this target.

To understand the impact of these gaps, we simulate a set of light curves for a single observation of WASP-127\,b and compare the errors on the transit depths when gaps are induced. Again we base the atmosphere off of current observations which suggest a large water abundance and potentially the presence of FeH \citep{skaf_w127}, which we model using the line lists from \cite{dulick_FeH,wende_FeH}.

The results of these fittings are shown in Figure \ref{fig:wasp127_errors}. The full light curve again has a wavelength dependent variation from the predicted radiometric errors but this is again relatively small. As expected, the fitting of the partial light curve results in larger uncertainties on the transit depth. In the case modelled, LC1 from Figure \ref{fig:w127_transits}, Twinkle only observes the target for 46\% of the transit. Using Equations 5 and 8, one would expect the error to increase by 48\% ($\sigma_p = \sigma_f \times \frac{1}{\sqrt{0.46}} = 1.476 \sigma_f$). We see the increase is wavelength dependent and generally between 20-40\%, less than predicted. Therefore the radiometric model may not always be capable of providing accurate error estimations. 

The recovered precision on different parameters is likely to be dependent upon the location of the gaps in the light curve. In this case the central portion of the transit is well sampled allowing for a precise recovery of the transit depth. However, ingress/egress are less well sampled and thus orbital parameters such as the inclination (i) and reduced semi-major axis (a/R$_*$) may be less well determined.

Furthermore, the standard methodology of analysing transiting exoplanet data is to fit to the light curves for planet-to-star radius ratio (R$_p$/R$_s$) to achieve a spectrum with error bars before performing atmospheric retrievals on said spectrum. This approach, which has essentially been followed here, distils time-domain observations down to a single point and thus much information about the orbital parameters of the system are lost. Fitting of full light curves (no gaps) usually retrieves the orbital parameters accurately but, as discussed, gaps can lead to less certainty. This potential degeneracy is lost in the standard method and so, to bring the data analysis one step closer to the raw data, retrievals with Terminus generated data could be conducted using the light curves themselves and the methodology described in \cite{yip}. The so called ``L-retrieval" allows for the orbital parameters (e.g. inclination, semi-major axis) to be free parameters in the retrieval to ensure that orbital degeneracies are accounted for. Such a methodology would be useful in the exploration of the effects of Earth obscuration, particularly as these orbital elements have been shown to be important in recovering the correct optical slope \citep{alexoudi_inc}. A thorough analysis is needed to explore this fully and Terminus can feed vital information into such an effort.

\section{Availability of Solar System Bodies}

Twinkle will also conduct spectroscopy of objects within our Solar System with perhaps the most promising use of the mission in this regard being the characterisation of small bodies. In particular, a diverse array of shapes for the 3 $\mu$m hydration feature, which generally cannot be observed from the ground, have been seen and used to classify asteroids \citep[e.g.][]{mastrapa,campins_nat,rivkin_emery,takir_emery,takir}. Twinkle's broad wavelength coverage will allow for studies of this spectral feature, and many others, as outlined in \citet{edwards_ast}.

\begin{figure}
    \centering
    \includegraphics[width = \columnwidth]{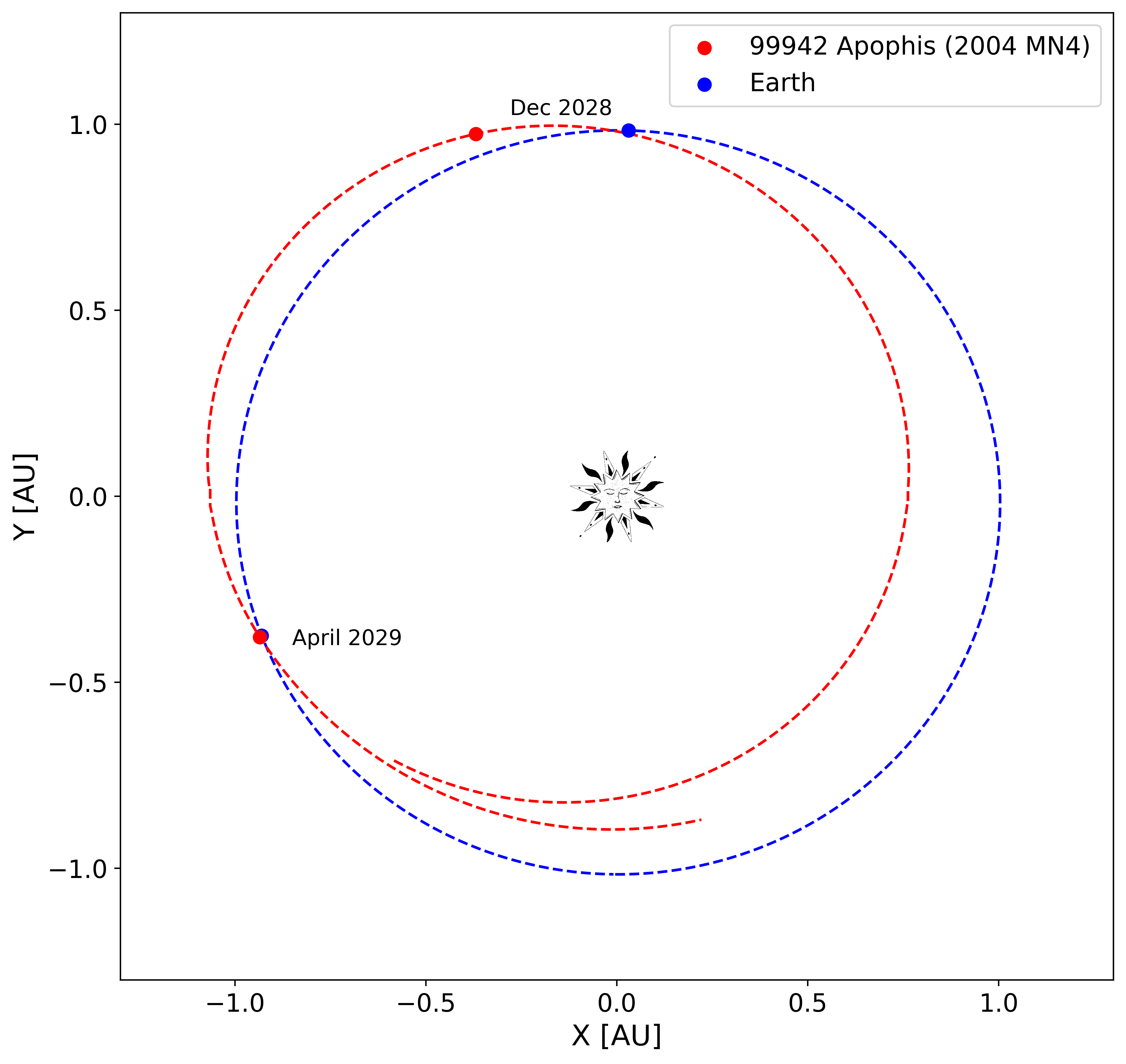}
    \includegraphics[width = \columnwidth]{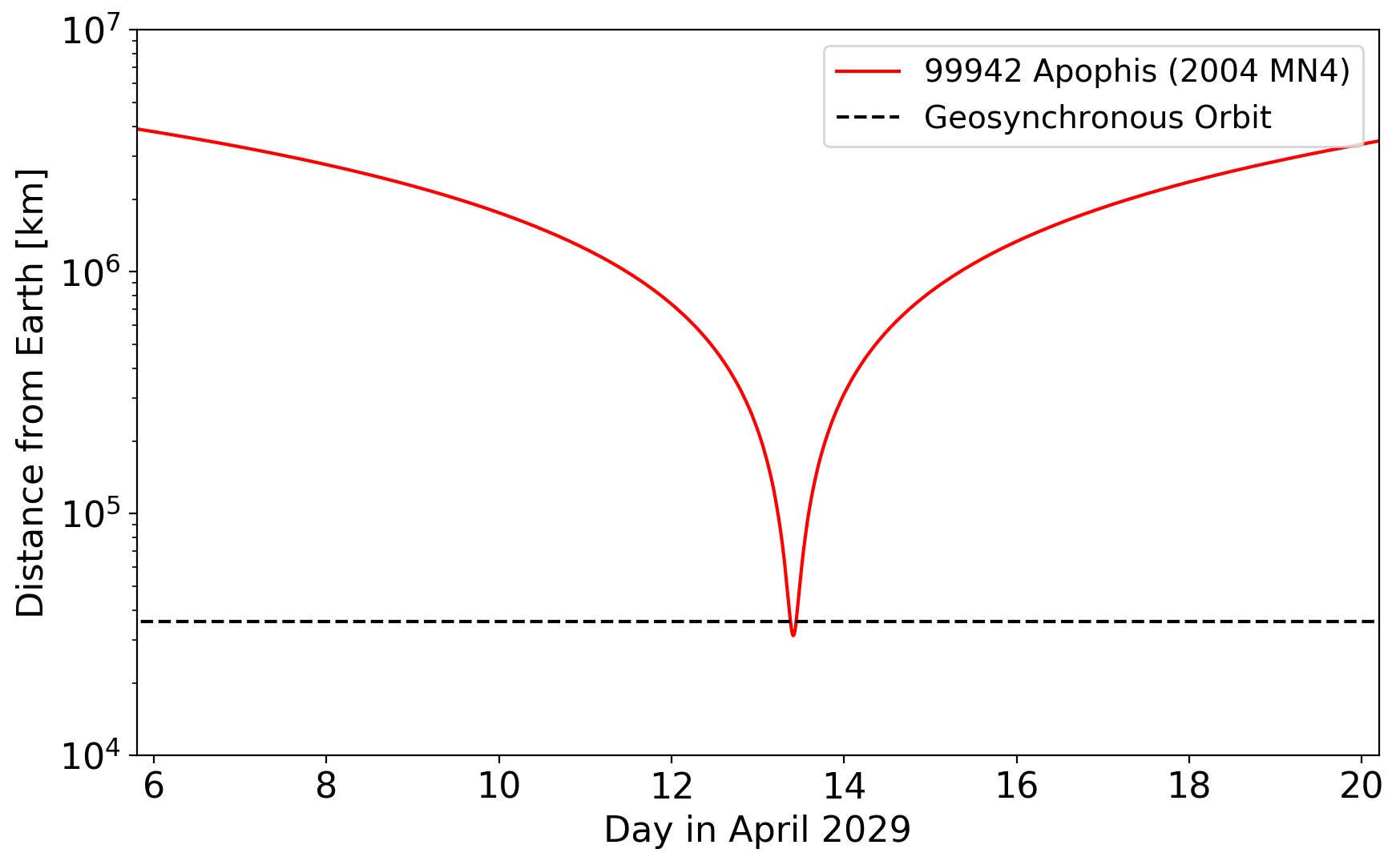}
    \caption{Top: orbit of Earth and Apophis from June 2028 to June 2029. In the period, Apophis crosses the orbit of Earth twice with the second of these crosses occurring during April 2029. Bottom: the distance between Earth and Apophis during the April 2029, highlighting that the minimum separation from the Earth surface is closer than geosynchronous satellites. Data for these plots was acquired via the NASA JPL Horizons service.}
    \label{fig:apophis_dist}
\end{figure}

The times at which major and minor Solar System bodies are within Twinkle's field of regard has previously been studied in \citet{edwards_ast, edwards_solar}. These studies showed that the outer planets, and main belt asteroids, will have long, regular periods within Twinkle's field of regard. However, the observation periods of Near-Earth Objects (NEOs) and Near-Earth Asteroids (NEAs) are far more sporadic. Hence, we revisit this analysis with the addition of considering Earth obscuration. For our example target, we choose 99942 Apophis (2004 MN4), a potentially hazardous asteroid (PHA). Apophis has a diameter of around 400 m \citep{lincadro_apophis,muller_apophis} and will have a close fly-by in 2029 (Figure \ref{fig:apophis_dist}). While it had been thought there was potentially a high probability of impact during this fly-by, or one in 2036, this has now been significantly downgraded \citep{krolikowska_apophis, chesley_apophis, thuillot_apophis}. Nevertheless, passing around 31,000\,km from the Earth's surface, Apophis will come within the orbits of geosynchronous satellites (see Figure \ref{fig:apophis_dist}). 

By comparing the data to likely meteorite analogues, current spectral analyses of Apophis have concluded it is an Sq-class asteroid that closely resembles the LL ordinary chondrite meteorites in terms of composition \citep{binzel_apophis, reddy_apophis}. This data was measured over 0.55-2.45 $\mu m$ and similarities have been noted to that of the asteroid Itokawa which was visited and studied by the Hayabusa mission \citep{abel_itokawa}.

\begin{figure*}
    \centering
    \includegraphics[width = \columnwidth]{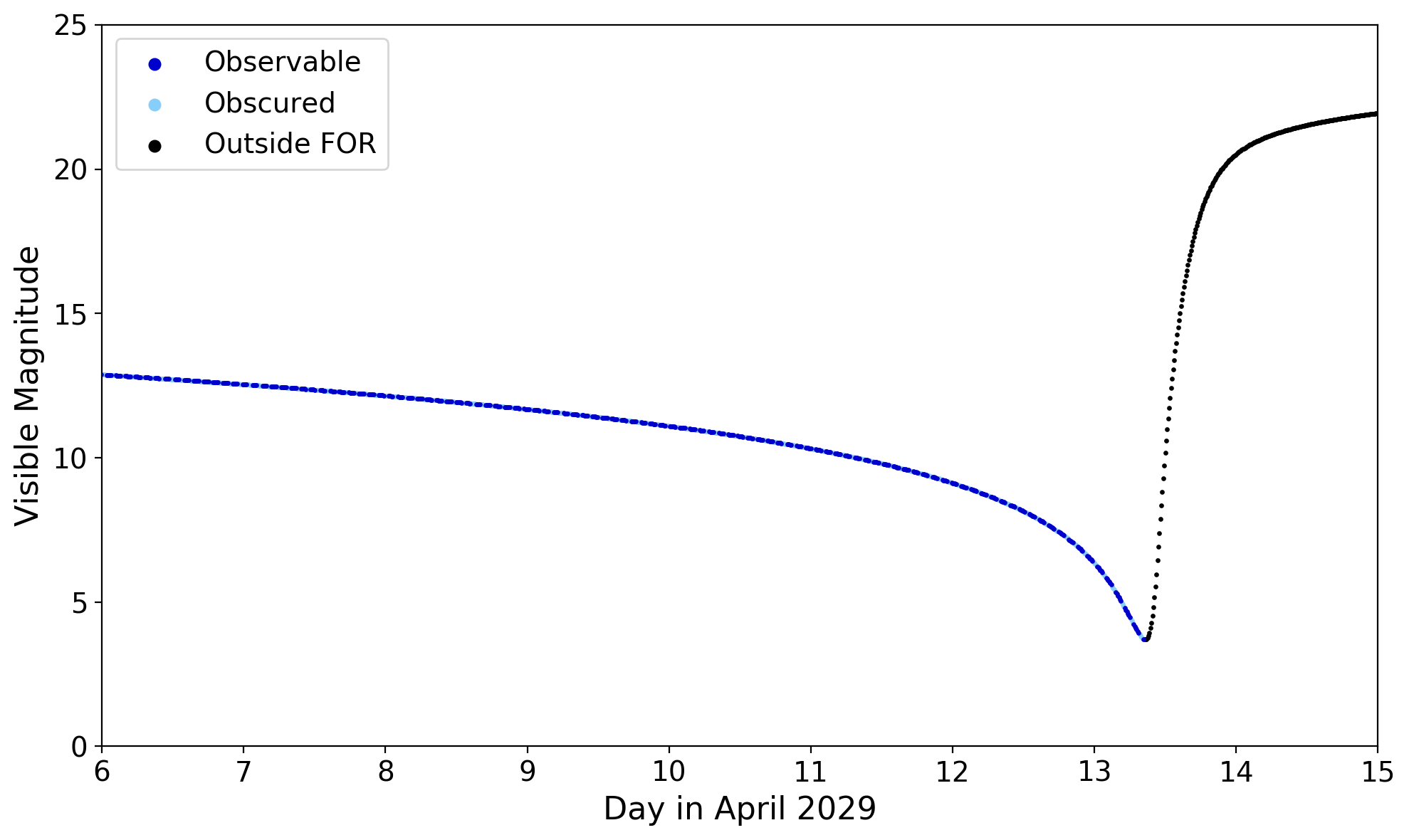}
    \includegraphics[width = \columnwidth]{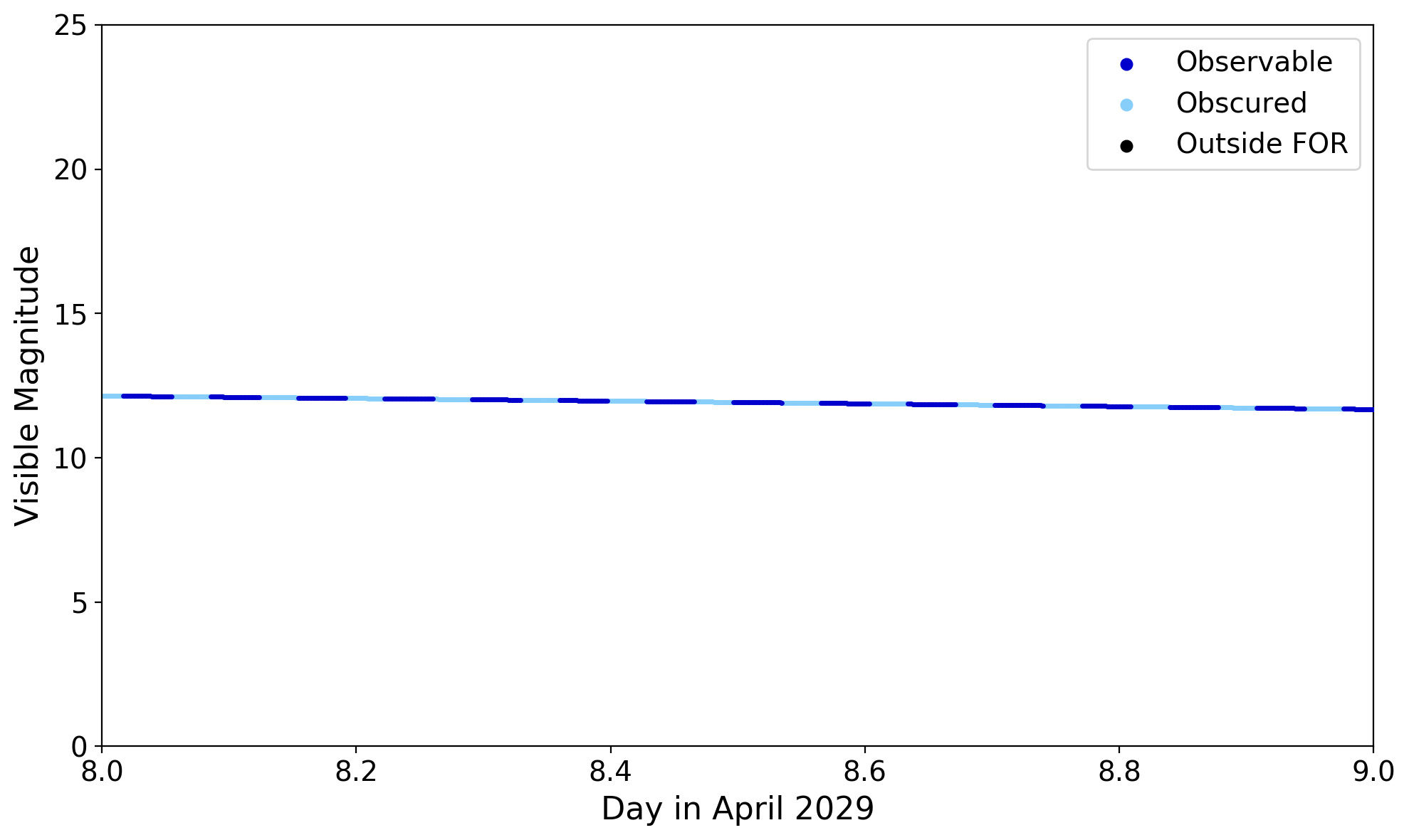}
    \includegraphics[width = \columnwidth]{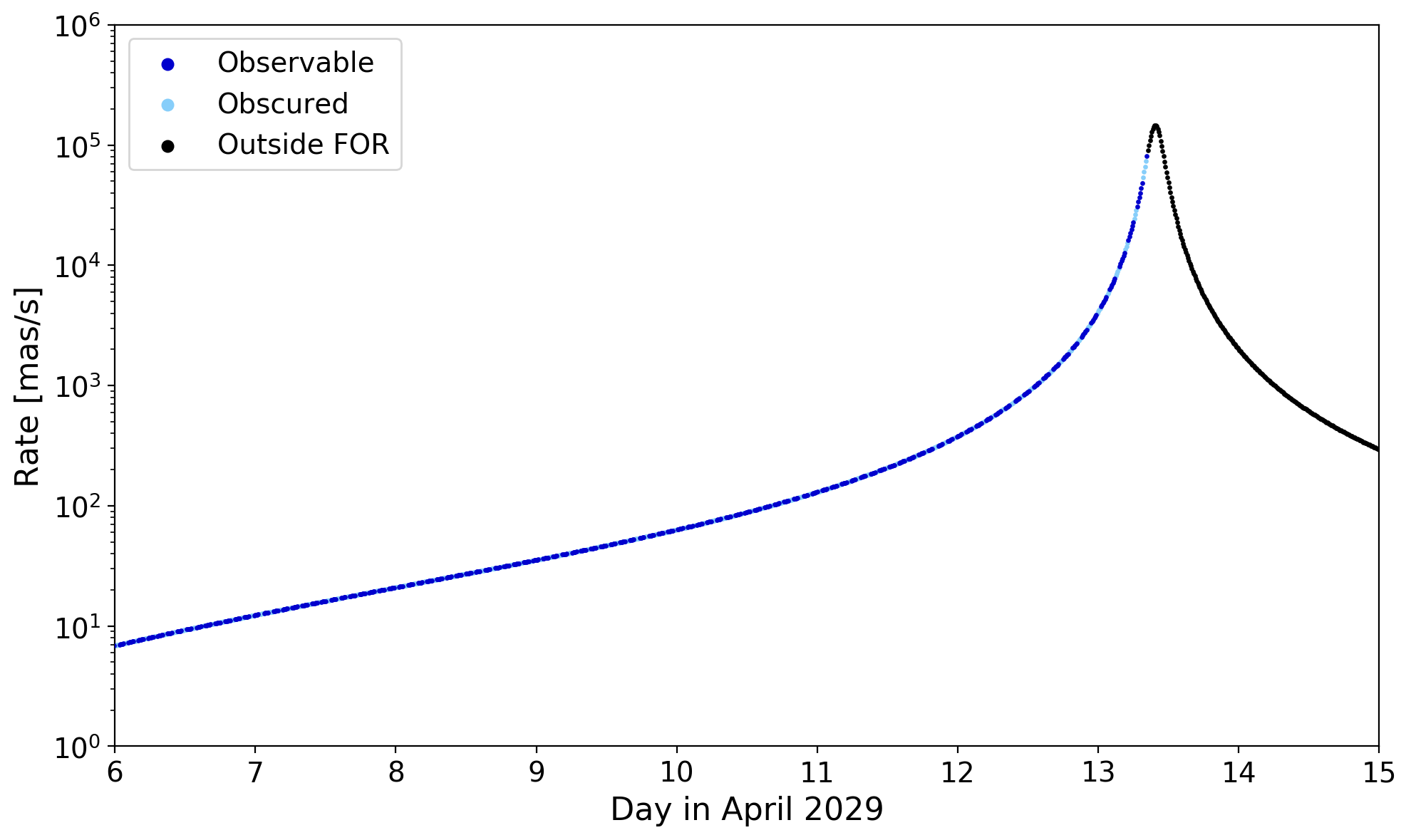}
    \includegraphics[width = \columnwidth]{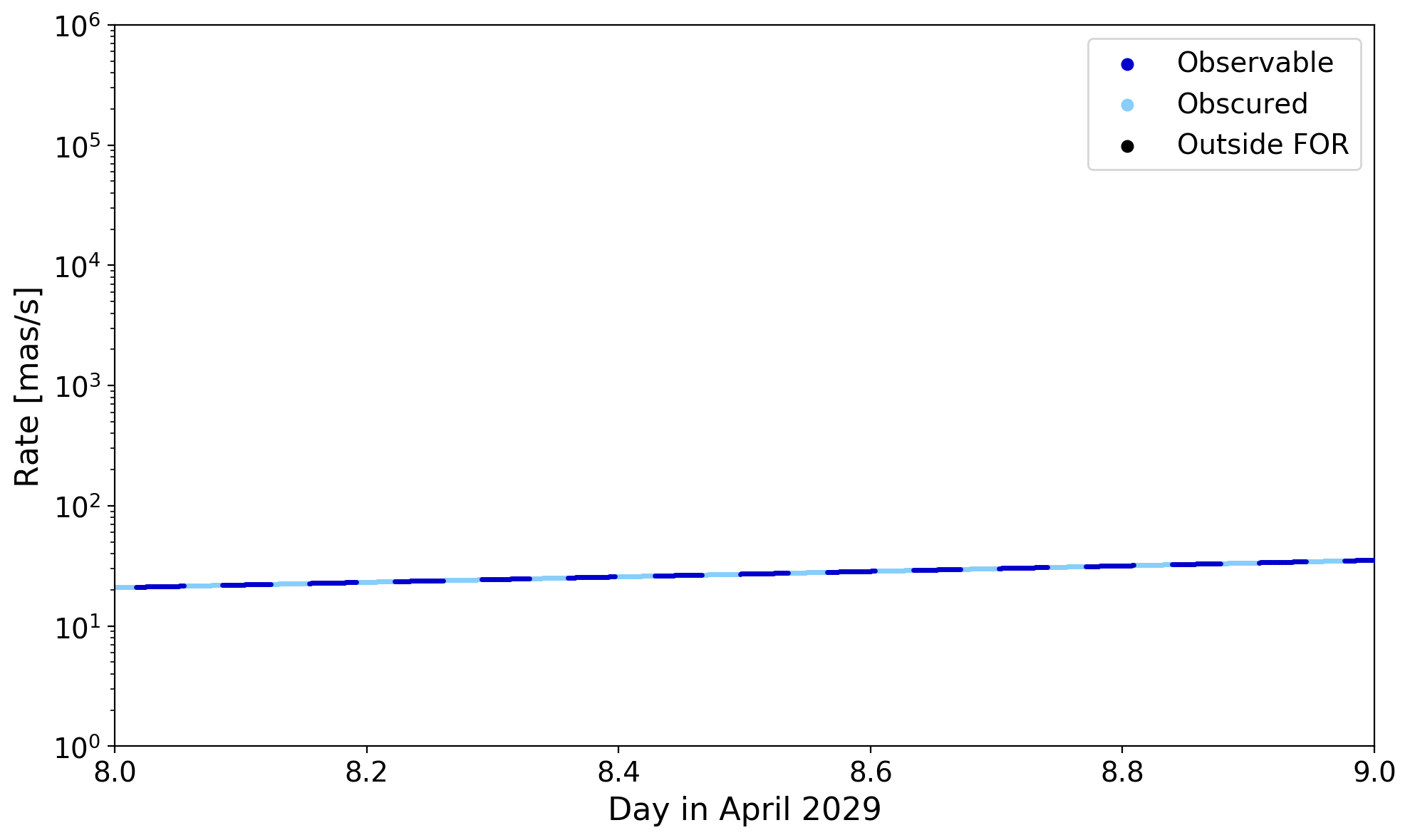}
    \caption{Visible magnitude (top) and rate of apparent motion (bottom) for Apophis during it's close fly-by in 2029. The availability of Apophis was checked at a cadence of 1 minute with dark blue indicating it is unobstructed, light blue showing times at which the Earth is occulting the target and black representing times when it has left the field of regard (i.e. exclusion due to Sun-target angle). The left-hand plots show these values for the week before the closest approach while the right-hand plots display the Earth obscuration more readily as Apophis approaches a rate of 30 mas/s.}
    \label{fig:apophis_movement}
\end{figure*}


\begin{figure*}
    \centering
    \includegraphics[width = 0.75\textwidth]{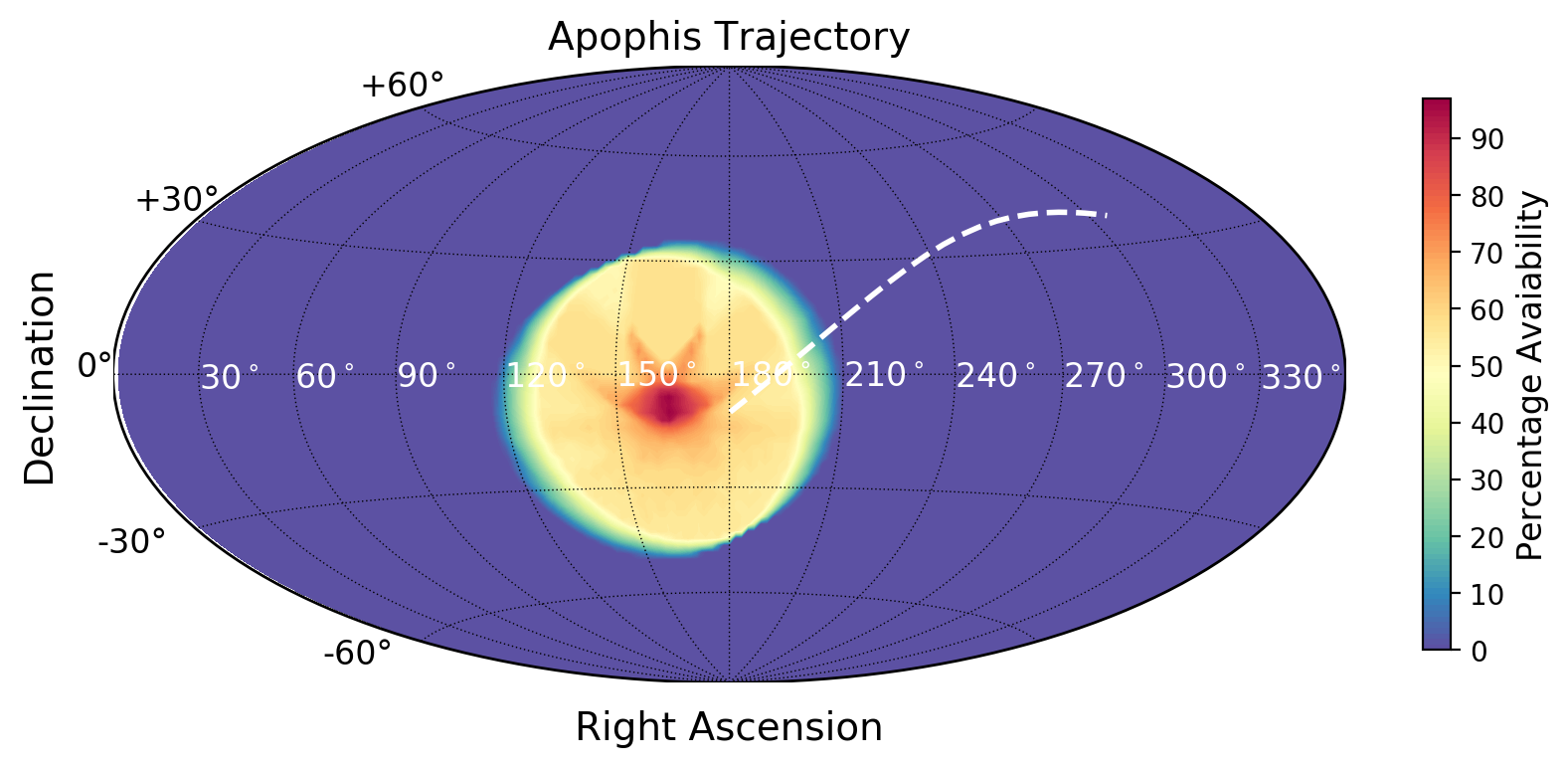}
    \caption{Average sky coverage during the two weeks before the closest approach of Apophis and the sky location of Apophis over that same period (white). It should be noted that, for the plotted Apophis trajectory, the time spent outside the FOR is only a few hours whereas the time spent within it equates to several days.}
    \label{fig:ast_path}
\end{figure*}

\begin{figure}
    \centering
    \includegraphics[width = \columnwidth]{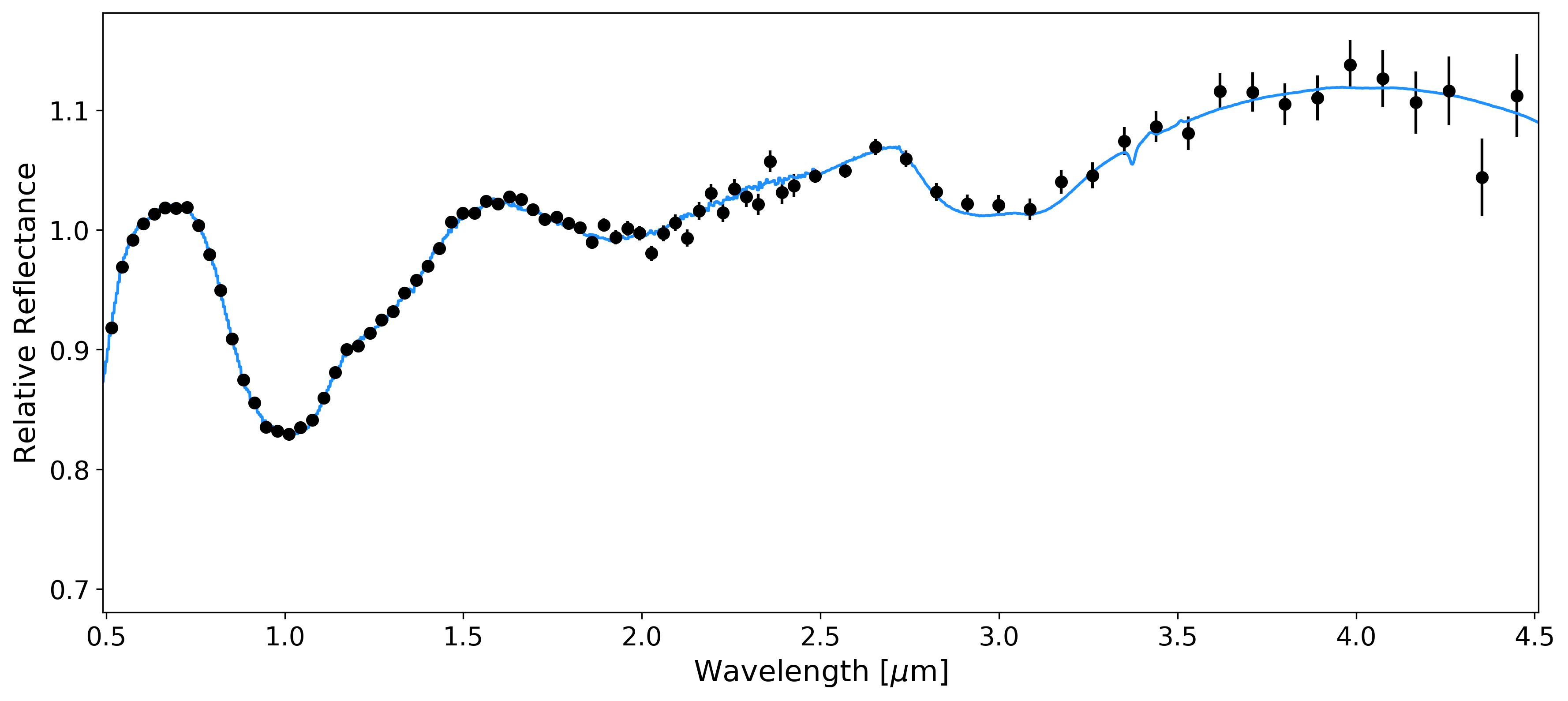}
    \caption{Simulated spectra for Apophis. The error bars are for a single exposure with a 300 s integration time on an object at a visible magnitude of 12. The spectrum is of an LL6 ordinary chondrite meteorite, taken from the RELAB database (bkr1dp015). We note that the reflectance shown here at shorter wavelengths ($<0.8 \mu m$) is slightly larger than found in actual studies of Apophis \citep{binzel_apophis, reddy_apophis}.}
    \label{fig:apophis_spec} 
\end{figure}

Here, we analyse the availability of Apophis over the week before, and day after, its closest approach to Earth. Terminus obtains asteroid ephemerides using the astropy API to the JPL Horizons database \citep{astropy}. In Figure \ref{fig:apophis_movement} we show the visible magnitude and apparent rate of motion of Apophis during this period. The interlaced dark and light blue segments show the availability of the asteroid before it leaves the field of regard soon after the closest point of its fly-by. The trajectory across the sky of Apophis is depicted in Figure \ref{fig:ast_path} along with the sky coverage of Twinkle over this period.

The ability of spacecraft to accurately track non-sidereal objects is key for their observation. The Spitzer Space Telescope was used extensively for characterising small bodies \citep[e.g.][]{trilling_cat,barucci} and tracked objects moving at rates of 543 mas/s \citep{trilling_exploreneo}. Spitzer was oriented using a inertial reference unit comprising of several high performance star trackers and observed asteroids using linear track segments. These were commanded as a vector rate in J2000 coordinates, passing through a specified RA and Dec at a specified time. The coordinates of the target can be obtained from services such as Jet Propulsion Laboratory’s Horizons System. JWST is expected be able to track objects moving at up to 30 mas/s \citep{thomas_jwst}.

The maximum rate at which Twinkle can track non-sidereal objects is still under definition but will be \textgreater 30 mas/s which we take here as a conservative maximum value. When this threshold is crossed, Apophis will have a visible magnitude of approximately 11.8. During the day or so before this rate limit is crossed, Apophis would be available for periods of 55 minutes, with 40 minute interruptions, again assuming a 20$^\circ$ Earth exclusion angle.


As demonstrated in Figure \ref{fig:apophis_spec}, such observation windows provide plenty of time to achieve high quality spectra. Here we simulated spectra for Apophis at a visible magnitude of 12 and an integration time of 5 minutes. We note that the thermal emission from the asteroid has been subtracted, which was modelled as a blackbody with a temperature of 300\,K, to give the relative reflectance of the asteroid. The input spectrum was taken from the RELAB database\footnote{\url{http://www.planetary.brown.edu/relab/}} and is of an LL6 ordinary chondrite meteorite.

Simulations have suggested the 2029 close encounter could cause landslides on Apophis, if the structure of some parts of the structure are significantly weak \citep{yu_apophis_sim}. The potential for resurfacing NEOs during terrestrial encounters in discussed in e.g. \cite{binzel_surf, nesvorny_surf} and spectral measurements can inform us on the freshness of the asteroid's surface, providing evidence for such mechanisms. Additionally, while an impact in 2029 has been ruled out, the potential for a future collision cannot be disregarded and further study of the object is needed to refine this. In particularly, the Yarkovsky effect has been shown to significantly alter predictions beyond 2029 and is sensitive to the physical parameters of Apophis, such as its albedo, diameter and density \citep{farnocchia_apophis,yu_apophis}. 

By observing Apophis simultaneously from 0.5-4.5\,$\mu m$, Twinkle could significantly inform the debate surrounding the nature of Apophis and it's potential threat level to Earth. Therefore, Twinkle could have a role to play in characterising known NEOs and NEAs, along with those predicted to be discovered by Near-Earth Object Surveillance Mission (NEOSM, \citet{mainzer_neocam}) and Vera C. Rubin Observatory, previously known as the Large Synoptic Survey Telescope (LSST, \cite{jones_lsst}). The ability of Twinkle to contribute to the study of NEOs and NEAs, and other specific asteroid populations, will be thoroughly detailed in further work.

\section{Conclusions and Future Work}

Terminus, a simulator with some time-domain capabilities has been developed to model observations with space-based telescopes. This model is especially applicable to exoplanets and can incorporate gaps in the light curve, caused by Earth obscuration, and be used to predict the potential impact on the accuracy of the retrieved atmospheric composition. Here, Terminus is baselined on the Twinkle Space Telescope but the model can be adapted for any space-based telescope and is especially applicable to those in a low-Earth orbit. 


The impact of gaps in exoplanet observations has not been fully explored and further work is needed. Obtaining a full transit, or eclipse, light curve is obviously the ideal case but when it is not possible, such as for HD\,209458\,b, an optimisation of the location, and length, of the gaps is required. By being able model when these gaps occur, it should be possible to begin to explore this by running multiple fittings and comparing the retrieved transit depth and atmospheric parameters. 


The Earth exclusion angle considered here is identical for the lit and unlit portions of the Earth. However, each will contribute different amounts of stray light and thus likely have separate exclusion angles. Future work will incorporate this capability, along with the capacity to quantitatively model the expected stray light from the Earth and Moon to firmly establish the exclusion angles required. The effect of different orbital parameters (e.g. altitude, 6am vs 6pm RAAN) can also be explored. Terminus will be updated to include the South Atlantic Anomaly (SAA) to model the impact in the event that the spacecraft must limit scientific operations during its ingress into this region. Other additional development aspects include satellite ground stations and calculating potential accesses to these facilities. Such capabilities will allow for the tool to serve wider concept of operations (CONOPS) concerns and, in the event that spacecraft design for any reason limits operations during downlink, this can then be accounted for in the scheduling. Additionally, Terminus could also be used to model other effects such as stellar variability or detector ramps such as those seen on Hubble and Spitzer.

Finally, Terminus will be incorporated into a web interface to provide the community with simulations of Twinkle's capabilities. Doing so will allow the tool to be more widely used and facilitate in-depth studies of Twinkle's capabilities. These could include modelling various atmospheric scenarios for each planet to judge its suitability for characterisation \citep[e.g.][]{fortenbach_jwst_pop}, performing retrievals on populations of exoplanets \citep[e.g.][]{changeat_alfnoor}, classifying groups of planets via colour-magnitude diagrams \citep[e.g.][]{dransfield_colour_mag}, testing machine-learning techniques for atmospheric retrieval \citep[e.g.][]{marquez_ml,zingales_exogan,hayes_ml,yip_blackbox} or the exploration of potential biases in current data analysis techniques \citep[e.g.][]{feng_bias,rocchetto_bias,changeat_2l,caldas_bias,powell_clouds,macdonald_bias,taylor_bias}. Additionally, thorough analyses of Twinkle's capabilities for specific scientific endeavours, such as confirming/refuting the presence of thermal inversions and identifying optical absorbers in ultra-hot Jupiters \cite[e.g.][]{fortney_2008,spiegel,haynes_w33,evans_wasp121_t2,parmentier2018,von_essen_w76, ares1,ares3,changeat_k9}, searching for an exoplanet mass-metallicity trend \citep[e.g.][]{wakeford_met,welbanks_met}, probing the atmospheres of planets in/close to the radius valley to discern their true nature \citep[e.g.][]{owen_ev,fulton,zeng_ww}, refining basic planetary and orbital characteristics \cite[e.g.][]{berado_spitzer,dalba,livingston_spitzer}, measuring planet masses through accurate transit timings \cite[e.g.][]{hadden_ttv,grimm_trappist,petigura_ttv}, verifying additional planets within systems \cite[e.g.][]{gillon,bonfanti_cheops}, studying non-transiting planets by measuring the planetary infrared excess \citep{stevenson_pie}, or even contributing to the search for exomoon candidates \citep[e.g.][]{simon_cheops_exomoon,heller_exomoon,teachey_exomoon}, can also be undertaken. \\



\section{Acknowledgements}

This work has utilised data from FreeFlyer, a mission design, analysis and operation software created by a.i. solutions. We thank Giovanna Tinetti, Marcell Tessenyi, Giorgio Savini, Subhajit Sarkar, Enzo Pascale, Angelos Tsiaras, Philip Windred, Andy Rivkin, Lorenzo Mugnai, Kai Hou Yip, Ahmed Al-Refaie, Quentin Changeat and Lara Ainsman for their guidance, comments and useful discussions. This work has been partially funded by the STFC grant ST/T001836/1.

\vspace{3mm}
\textbf{Software:} TauREx3 \citep{al-refaie_taurex3}, pylightcurve \citep{tsiaras_plc}, ExoTETHyS \citep{morello_exotethys}, ExoSim \citep{Exosim_2}, Astropy \citep{astropy}, h5py \citep{hdf5_collette}, emcee \citep{emcee}, Matplotlib \citep{Hunter_matplotlib}, Multinest \citep{Feroz_multinest,multinest}, Pandas \citep{mckinney_pandas}, Numpy \citep{oliphant_numpy}, SciPy \citep{scipy}, corner \citep{corner}.

\bibliographystyle{aasjournal}
\bibliography{main}
\end{document}